\documentclass[fleqn,usenatbib]{mnras}

\usepackage{newtxtext,newtxmath}
\usepackage{mathtools}

\usepackage[T1]{fontenc}

\DeclareRobustCommand{\VAN}[3]{#2}
\let\VANthebibliography\thebibliography
\def\thebibliography{\DeclareRobustCommand{\VAN}[3]{##3}\VANthebibliography}

\usepackage{graphicx}	
\usepackage{amsmath}	
\usepackage{gensymb}
\usepackage{hyperref}

\title[Mechanism-independent modeling of pulsars]{A mechanism-independent methodology for modeling $\gamma$-ray phaseograms of pulsars in the framework of north-south symmetry}

\author[P. K. H. Yeung et al.]{
Paul K. H. Yeung,$^{1}$\thanks{E-mail: pkh91yg@icrr.u-tokyo.ac.jp (PKHY)}
Dmitry Khangulyan,$^{2,3}$
and Takayuki Saito$^{1}$
\\
$^{1}$Institute for Cosmic Ray Research, University of Tokyo, 5-1-5, Kashiwa-no-ha, Kashiwa, Chiba 277-8582, Japan\\
$^{2}$Key Laboratory of Particle Astrophyics, Institute of High Energy Physics, Chinese Academy of Sciences, 100049 Beijing, China\\
$^{3}$Tianfu Cosmic Ray Research Center, 610000 Chengdu, Sichuan, China
}

\date{Accepted 2025 July 9. Received 2025 July 9; in original form 2025 April 8}

\pubyear{\the\year{}}

\newcommand{\PKHY}[1]{{{#1}}}
\newcommand{\REFE}[1]{{{#1}}}
\newcommand{\REFESEC}[1]{{{#1}}}

\begin{document}
\label{firstpage}
\pagerange{\pageref{firstpage}--\pageref{lastpage}}
\maketitle

\begin{abstract}

Fermi-LAT observations revealed that each GeV phase-folded light-curve (aka. phaseogram) of the Crab, Geminga, Dragonfly and Vela pulsars consists of two pulses (P1 \& P2) and a “Bridge” between them. There is clearly a “bump” at the Bridge phase of Vela’s pulse profiles, that could also be regarded as the third pulse (P3). Differently, the Crab’s, Geminga’s \& Dragonfly’s Bridges relatively resemble a “valley floor”. Despite such an apparent difference, it is interesting to investigate whether their Bridge emissions are still within the same general picture as Vela’s. Assuming the north-south symmetry, we would expect the fourth component (Bridge2/P4) to exist as well. However, such a hypothetical Bridge2/P4 is not intuitively identified on $\gamma$-ray phaseograms of the Crab, Geminga, Dragonfly and Vela pulsars. It is also intriguing to hint at the rationale for the non-discovery of Bridge2/P4. Our prototypical toy model is free of assumptions on emission regions or radiation mechanisms. Instead, it assumes a north-south symmetric geometry and one circularly symmetric beam per hemisphere, while taking into account  
\PKHY{Doppler shifts (the most innovative element)}, time delays and  \PKHY{energy-dependent beam shapes}. Tentative compatibility of our fitting results with wind models is reported. Notably, for the Crab pulsar, we found a preliminary qualitative correlation between our model predictions and the IXPE results on X-ray polarisation. The softer $\gamma$-ray pulsation of the Geminga pulsar is found to span over its full phase. Prompted by  systematic evaluations, we outline some potential improvements for our toy model.

\end{abstract}




\section{Introduction}

Following the discovery of pulsars by \citet{1968Natur.217..709H},
they were rapidly associated with neutron stars emitting beams of
radiation along magnetic field lines misaligned with the rotational axis
\citep{1968Natur.218..731G}. Shortly after putting forward this
hypothesis, it was shown that there are fundamental differences
between the structure of pulsar magnetic field and the
idealised case of magnetised star in vacuum
\citep{1955AnAp...18....1D}.  \citet{1969ApJ...157..869G} have shown that
the rotation of a magnetised neutron star induces strong electric
fields that extract charges from the stellar surface, creating a
dense, co-rotating plasma. Their model introduced the concept of the
Goldreich-Julian charge density and the light cylinder, delineating
the boundary between closed and open magnetic field
lines. Furthermore, \citet{1971ApJ...164..529S} showed that the pair
creation process operates very efficiently in the magnetosphere, providing
an ample and distributed source of charges.

\REFE{One can perform sufficiently deep surveys in the radio and $\gamma$-ray bands to establish statistically significant populations of radio and $\gamma$-ray pulsars. While their observational manifestations share some similarities, the two populations differ markedly due to their distinct radiation mechanisms and the capabilities of instruments operating in these bands. $\gamma$-ray instruments typically have modest effective area, so they primarily detect pulsars with high spin-down luminosities. Moreover, generating $\gamma$-ray emission appears to require faster rotation, so the $\gamma$-ray pulsar population is dominated by young pulsars and recycled millisecond pulsars. In contrast, radio emission can be detected from older, less energetic pulsars. $\gamma$-ray emission is believed to arise in the outer magnetosphere, forming broader emission beams. This increases the likelihood that our line of sight (l.o.s.) intersects a $\gamma$-ray beam, enhancing its detectability even if radio emission is not detected. On the other hand, radio emission is thought to be narrower beams, that are originated near the pulsar surface and aligned with open magnetic field lines. This geometrically restricts the detectability of radio beams.}

A key development in understanding the origin of pulsed emission came
with the recognition that a strong electric field exists in regions
where the local charge density cannot maintain the Goldreich-Julian
value. \citet{1971ApJ...164..529S} proposed that particles accelerated
in these regions emit beams of non-thermal photons. \REFE{Owing to the synchronised rotation of a pulsar and its emission sites}, these beams manifest
themselves as pulsed emission. This seminal idea was elaborated into
 the polar cap \citep{1975ApJ...196...51R}, slot gap
\citep{1979ApJ...231..854A} and outer gap models
\citep{1986ApJ...300..500C}.

These fundamental results show that the locations of the production
sites of the pulsed emissions remain uncertain, and geometrical
factors play an essential role in the formation of pulsars’ phase-folded light-curves (aka. phaseograms).
When a pulsar is self-rotating rapidly, the trajectory of our l.o.s. is cutting through different parts of each pulsar beam periodically. This is the reason why we detect periodic emission from pulsars. Under an assumption of a circular, centrally peaked beam, when our l.o.s. is at its closest approach to the beam axis (i.e. the central direction of the beam), the received count rate roughly approaches a peak. Fundamentally speaking, a beam’s “real” peak is never observable, since there is always a non-zero angular separation between our l.o.s. and the beam axis.

\REFE{The Crab pulsar (PSR J0534+2200), Geminga pulsar (PSR J0633+1746), Dragonfly pulsar (PSR J2021+3651) and Vela pulsar (PSR J0835-4510)} are among the brightest $\gamma$-ray pulsars detected by \emph{Fermi} Large Area Telescope (LAT) so far \citep[LAT Third Catalog of Gamma-ray Pulsars;][]{Smith_3PC_2023}. Each of their  phaseograms demonstrates double asymmetric pulses (P1 \& P2) and an inter-pulse Bridge.  There is not only asymmetry between the leading wing (LW) and trailing wing (TW) of a pulse, but also asymmetry between P1 and P2 of a pulsar. Nevertheless, we emphasise that these asymmetric features shown on phaseograms do not necessarily imply asymmetric beam shapes or north-south asymmetry. Theoretically, even if a pulsar has a north-south symmetric geometry and emits circularly symmetric beams, some geometrical and physical effects could coalitionally deform its pulse profile to the asymmetric form that we detect. Our modelings presented in this article are within such a framework.

Unlike the valley-floor-like Bridges of the Crab, Geminga and Dragonfly pulsars \citep[e.g.][]{Fierro_EGRET_1998,  Abdo_Crab_2010, Abdo_GemingaPSR_2010, Aleksic_BD_2014, Ahnen2016, MAGIC_GemingaPSR_2020, Yeung_CrabPSR_2020, CTAO-LST_CrabPSR_2024, CTAO-LST_GemingaPSR_2025, Wang_thesis_2025}, the Bridge of the Vela pulsar has an arch-like structure \citep[i.e. a “bump” of the detected count rate; e.g.][]{Fierro_EGRET_1998,Abdo2010_VelaPSR,Leung2014,HESS_Vela_2023, Kargaltsev2023}. Accordingly, the Vela pulsar’s Bridge is also called a third pulse (P3). Despite this apparent difference, one of our goals is to investigate whether the Bridges/P3 of these four pulsars could be within the same general picture.

Under our assumption of the north-south symmetry, everything in the environment of a pulsar is expected to exist in pairs. For instance, the origins of P1 and P2 would be the mirror of each other, at the northern and southern hemispheres of the pulsar respectively. Likewise, we would expect the origin of the Bridge/P3 to have a counterpart emitting a fourth component “Bridge2/P4” at the opposite hemisphere. However, such a hypothetical “Bridge2/P4” is not identified on any $\gamma$-ray phaseogram of Crab, Geminga, Dragonfly and Vela pulsars. Another of our goals is to hint at the rationale for the non-discovery of “Bridge2/P4”. One possibility is that “Bridge2/P4” is mixed up with and indistinguishable from other components, while another possibility is that our l.o.s. disfavours the detection of this hypothetical fourth component.

As demonstrated in some of the aforementioned literature, a $\gamma$-ray pulsar typically has an energy-dependent pulse shape and, equivalently, a phase-dependent spectral shape. These could be indirect implications that the broadband $\gamma$-ray emission of a pulsar is originated from multiple emission sites (outer gap, wind current sheet, etc.) and/or multiple radiation mechanisms (synchrotron/curvature radiation, inverse-Compton scattering, etc.), as put forward by \citet{Yeung_CrabPSR_2020} and \citet{Harding_model_2021}, for instance. For such hybrid-origin scenarios, the estimates of the transition energies among different components are somewhat sketchy and challenging.

This article advocates an innovative modeling methodology that is independent of emission sites and radiation mechanisms. We establish a prototypical toy model and test it on \emph{Fermi}-LAT $\gamma$-ray phaseograms of Crab, Geminga, Dragonfly and Vela pulsars. We highlight the major properties inferred with this model, evaluate its predictive power, and propose some improvements that will potentially reconcile the existing deficiencies.

\section{Methodology}

Our prototypical toy model is free of assumptions on emission regions or radiation mechanisms. Instead, we make assumptions on the emission geometry and beam shapes only. 

\subsection{Geometrical properties}

   \begin{figure}
   \centering
   \includegraphics[width=.99\columnwidth]{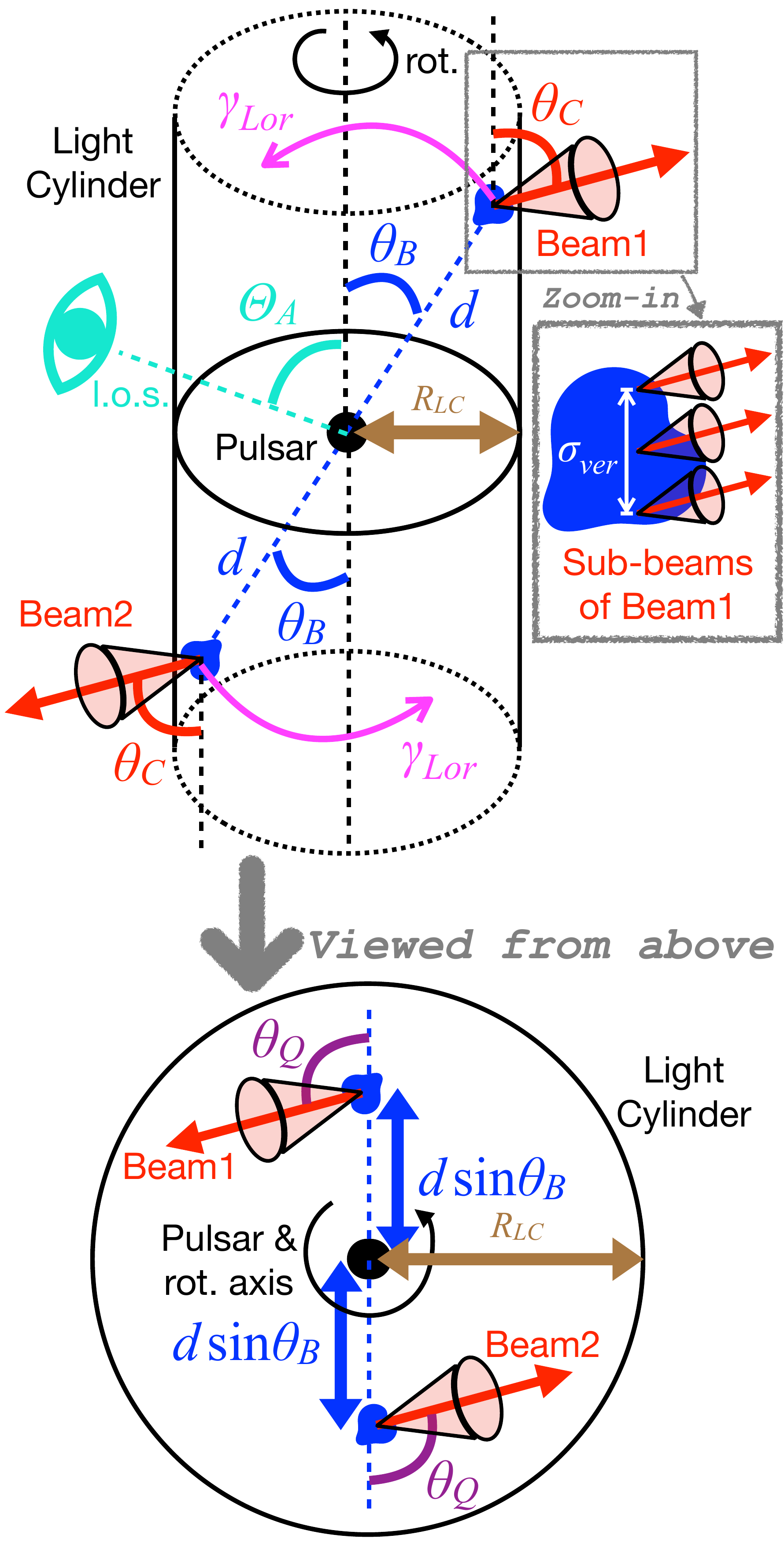}
   \caption{Schematic illustration of a north-south symmetric environment of a pulsar (side view and top view respectively). We use the blue and red colours to depict the emission sites and beams respectively. The zoom-in window provides additional elucidation on how we use a three-point approximation to characterise the vertical extension of an emission site. It is assumed that different sub-beams of the same beam are axis-aligned and share an identical distribution of emission count rates. Definitions of notations can be found in Table~\ref{ParDef}.
}
              \label{Schematic}%
    \end{figure}

Throughout this paper, we denote all lines parallel to the pulsar’s rotational axis as "vertical", and all planes parallel to the equatorial plane (i.e. normal to the rotational axis) as "horizontal". On a given horizontal plane, each point is specified by a “radial” coordinate (its shortest distance from the rotational axis) and an “azimuthal” coordinate (the angle between its position vector and a reference vector, where these two vectors intersect at the rotational axis). The equatorial plane separates a pulsar into northern and southern hemispheres.  We assume a north-south symmetric configuration (schematically demonstrated in Fig.~\ref{Schematic}), where everything at the north is exactly opposite to a counterpart at the south. Our another assumption is that  each hemisphere only harbours a single  emission site emitting a single circular beam.  The first parameter we define is the angle between our l.o.s. and the  rotational axis ($\Theta_A$).

Realistically, an emission site of a pulsar is an unknown three-dimensional structure. Our prototypical toy model determines its most representative location, which is parametrised as its separation from the pulsar center ($d$) and its zenith angle measured from the rotational axis ($\theta_B$). For the sake of flexibility in the fitting procedure, we transform $d$ and ${\theta_B}$ into the radial separation ($d\sin{\theta_B}$) and vertical separation ($d\cos{\theta_B}$).

We specify the direction of a beam axis by using two coordinates: the beam axis's zenith angle measured from the rotational axis ($\theta_C$), and the azimuthal angle between the beam axis and emission site ($\theta_Q$; values between $0\degree$ and $180\degree$ mean that the beam axis is ahead of the emission site). We name the northern and southern beams as Beam1 and Beam2 respectively.

The angle between the Beam1 axis and our l.o.s. ($\psi$) oscillates as a function of the time in the pulsar frame ($t_{psr}$):
\begin{equation}\label{GEOeqn}
    \psi=\cos^{-1}{[\cos{(2{\pi}t_{psr}+\theta_Q)}\sin{\theta_C}\sin{\Theta_A}-\cos{\theta_C}\cos{\Theta_A}]} \ \ \ .
\end{equation}
For any given $t_{psr}$, the Beam2 axis--l.o.s. angle is $180\degree-\psi$.

The magnetic pole of a pulsar is not parametrised in our toy model. Nonetheless, we speculate that its orientation is somehow related to other geometrical properties (e.g. emission site locations, beam axis directions). Hence, a qualitative constraint on the magnetic pole’s orientation might be derived from this model.

\subsection{Time delay}

Since the emission site is at a separation $d$ from the pulsar center, the lengths of times taken for north and south beams to reach us are different. Also,the photons’ travel times vary from phase to phase. As a result, the phase in our detector frame ($t_{det}$) changes non-linearly with $t_{psr}$. More importantly, the conversion from $t_{det}$ to $t_{psr}$ for Beam1 is different from that for Beam2:
\begin{equation}\label{TDeqn}
 \begin{aligned}
    &t_{det}-t_{psr}-k=\\ &\begin{cases} 
    \frac{d}{cT}(-\cos{2{\pi}t_{psr}}\sin{\theta_B}\sin{\Theta_A}+\cos{\theta_B}\cos{\Theta_A}) & \text{for Beam1}\\
    \frac{d}{cT}(\cos{2{\pi}t_{psr}}\sin{\theta_B}\sin{\Theta_A}-\cos{\theta_B}\cos{\Theta_A}) & \text{for Beam2}
    \end{cases} \ \ \ ,
 \end{aligned}
\end{equation}
where $c$ is the speed of light, $T$ is the rotational period of the pulsar and $k$ is a time-independent term related to the phase definition of a phaseogram. Here we define an additional free parameter $t_0$, which is the $t_{det}$ corresponding to $t_{psr}=0$ (i.e. the moment when the Beam1's emission site is at its closest approach to us), so that $k=t_0-\frac{d}{cT}\cos{(\Theta_A+\theta_B)}$. In this way, $t_0$ characterises the azimuthal location of the emission site at a specific phase.

The effect of time delay could partially account for the measured P2--P1 phase interval of < 0.5, as well as the observed asymmetry between the LW and TW of a pulse. We define $t_1$ and $t_2$ as the phases (in our detector frame) when the Beam1 and Beam2 axes are closest to our l.o.s., respectively. We use Eqn.~\ref{TDeqn} to derive an analytical relation among $t_1$, $t_2$, $\Theta_A$, $d$ and $\theta_B$:
\begin{equation}\label{t1_t2_intv}
    t_2 - t_1 = 0.5 - \frac{2d}{cT}\cos{\theta_B}\cos{\Theta_A} \ \ \ .
\end{equation}
Also, $t_1$ is related to $t_0$  as follows:
\begin{equation}
    t_1 = t_0 - \frac{\theta_Q}{360\degree} - \frac{d\sin{\theta_B}}{cT}(\cos\theta_Q-1)\sin{\Theta_A} \ \ \ .
\end{equation}

\subsection{Doppler shift \& radially extended emission region}

When a pulsar is rotating, each of its emission site moves towards us and away from us in countless cycles. As a result, the observed beam is redshifted and blueshifted alternately. Moreover, the level of redshift/blueshift varies from phase to phase. The most critical part of our model is that the photons with the same detected energy ($E_{det}$) at different phases correspond to different unshifted energies ($E_{psr}$). The Doppler effect further modulates the P2--P1 phase interval and the LW--TW asymmetry. It is very important to note that, because of the Doppler shift, the observed peak phases are somewhat different from $t_1$ and $t_2$ (defined in the previous subsection).

Inside the light cylinder of a pulsar,  every emission point moves coherently with the pulsar's rotation. Therefore, the linear speed ($v_i$) of an individual emission point is proportional to its radial separation ($d_i\sin{\theta_{Bi}}$) from the rotational axis: $v_i={2\pi}d_i\sin{\theta_{Bi}}/T$.
However, this equality does not hold true when it comes to the collective motion of a radially extended emission site\footnote{Mathematically, $\overline{d\sin{\theta_B}}\neq\overline{d}\sin{\overline{\theta_B}}$, where $\overline{d\sin{\theta_B}}$, $\overline{d}$ and $\overline{\theta_B}$ represent the average values of $d_i\sin{\theta_{Bi}}$, ${d_i}$ and ${\theta_{Bi}}$ respectively.}. 
Hence, we introduce an additional free parameter: the Lorentz factor $\gamma_{Lor}=\frac{1}{\sqrt{1-\frac{v^2}{c^2}}}$, where $v$ is the average linear speed of the collective motion of an emission site.
We substitute $\gamma_{Lor}$ and $v$ to compute the inverse Doppler factor ($\varepsilon$) as a function of $t_{psr}$. For Beam1,
\begin{equation}\label{DSeqn}
    \varepsilon=\frac{E_{psr}}{E_{det}}=\frac{1}{\gamma_{Lor}\left(1-\frac{v}{c}\sin{\Theta_A}\sin{2{\pi}t_{psr}}\right)} \ \ \ .
\end{equation}
Replacing $t_{psr}$ with $t_{psr}+0.5$ gives the expression for $\varepsilon$ of Beam2.

In this way, the radial extension scale ($\sigma_{rad}$)  could be roughly estimated by comparing the inferred $\gamma_{Lor}$, $d$ and $\theta_B$:
\begin{equation}\label{hor_ext}
    \sigma_{rad}=\frac{v}{c}R_{LC}-d\sin{\theta_B}=\sqrt{1-\frac{1}{\gamma_{Lor}^2}}R_{LC}-d\sin{\theta_B} \ \ \ ,
\end{equation}
where $R_{LC}$ is the light cylinder radius and is equal to $\frac{cT}{2\pi}$. For an emission region inside the light cylinder, the Lorentz factor of an individual point diverges to infinity if it approaches the light cylinder wall, so we would expect $\sigma_{rad}$ to be positive. If we obtain a negative $\sigma_{rad}$, this might imply that the emission site extends beyond the light cylinder and its particles outside move at relatively lower speeds, pulling down the overall average $\gamma_{Lor}$. Whereas, if $d\sin{\theta_B}$ of the most representative location exceeds $R_{LC}$, then the collective motion would be mostly incoherent with the pulsar's rotation (relevant clarifications are made in Appendix~\ref{ApxBeyondLC}). In such a case, Eqn.~\ref{hor_ext} always yields a negative $\sigma_{rad}$, that does not reflect the genuine radial extension at all.

In addition to energies of individual photons, Doppler shift also affects intensities in this way:
\begin{equation}\label{DSint}
    I_{det}=\frac{I_{psr}}{\varepsilon^3} \ \ \ ,
\end{equation}
where $I_{det}$ and $I_{psr}$ are the “specific intensities” in the detector frame and the pulsar frame respectively (detailed explanations are presented in Appendix~\ref{ApxIntensity}).


\begin{table*}
	\centering
	\caption{List of geometrical parameters and their definitions. }
	\label{ParDef}
	\begin{tabular}{cl} 
		\hline
		Notation & Definition \\
		\hline
 \multicolumn{2}{c}{\underline{Free geometrical parameters for iterations}}  \\
		$\Theta_A$ & The angle between our l.o.s. and the pulsar’s rotational axis. \\
		$d\sin{\theta_B}$ & The average radial separation of the emission site from the pulsar center.$^{\dag}$ \\
		$d\cos{\theta_B}$ & The average vertical separation of the emission site from the pulsar center.$^{\dag}$ \\
		$t_0$ & The phase (in the detector frame) corresponding to the moment when the Beam1's emission site is at its closest approach to \REFE{our l.o.s.}. \\
		$\theta_C$ & The beam axis's zenith angle measured from the rotational axis. \\
		$\theta_Q$ & The azimuthal angle by which the beam axis is ahead of the emission site. \\
		$\gamma_{Lor}$ & The average Lorentz factor of the collective motion of the emission site. \\
		$\sigma_{ver}$ & The characteristic vertical extension of the emission site.  \\
 \multicolumn{2}{c}{\underline{Additional geometrical properties}*}  \\
		$d$ & The average separation of the emission site from the pulsar center.$^{\dag}$ \\
		${\theta_B}$ & The emission site's average zenith angle measured from the rotational axis.$^{\dag}$ \\
		$t_1$ & The phase (in the detector frame) when the Beam1 axis is at its closest approach to our l.o.s..$^{\ddag}$ \\
		$t_2$ & The phase (in the detector frame) when the Beam2 axis is at its closest approach to our l.o.s..$^{\ddag}$ \\
		$\sigma_{rad}$ & The estimated radial extension scale of the emission site.$^{\diamond}$  \\
		\hline
	\end{tabular}\\
* These additional properties are computed by substituting some iterated parameters into specific equations. \\
$^{\dag}$ For the sake of flexibility in the iteration procedure, we transform $d$ and ${\theta_B}$ into $d\sin{\theta_B}$ and $d\cos{\theta_B}$. \\
$^{\ddag}$ It is slightly different from the actually observed peak phase, due to the modulation by the Doppler shift. \\
$^{\diamond}$ This estimation is inappropriate if $d\sin{\theta_B}$ exceeds $R_{LC}$.
\end{table*}

\subsection{Vertical extension of each emission site}

For a vertically extended emission region,  the detected pulse on a phaseogram could be regarded as an ensemble of sub-pulses emitted at different heights. Each sub-pulse and its counterpart at the opposite hemisphere have a different $t_2 - t_1$, which varies linearly with the height of its emission point (according to Eqn.~\ref{t1_t2_intv}).

We use a three-point approximation to characterise the vertical extension. To be specific, we define three points sharing the same radial and azimuthal coordinates but at equidistant heights. These three points are assumed to emit sub-beams with parallel axes and a shared count rate distribution, as schematically elucidated in the zoom-in window of Fig.~\ref{Schematic}. Hereby, we clarify that $d$, $\theta_B$ and $t_0$ are defined for the middle point, and we quantify the characteristic vertical extension as an extra variable: the separation between top and bottom points ($\sigma_{ver}$).

Taking into account the radial and vertical extensions but neglecting the azimuthal extension, our prototypical toy model approximates the three-dimensional structure of an emission region as a two-dimensional plane. We omit the \PKHY{location-dependent variations of sub-beam properties}.

\subsection{Distribution function of detected count rate}

Under our assumption of north-south symmetry, the emission count rates of Beam1 and Beam2 in the pulsar frame follow an exactly identical distribution. Regarding the received count rate ($C$) of each beam in our detector frame, we consider the $\psi$-dependence and $\varepsilon$-dependence\footnote{For each phaseogram, $\varepsilon$-dependence can be considered equivalent to $E_{psr}$-dependence, neglecting the interval of $E_{det}$.}, as well as the coupling between them. As aforementioned, we omit the \PKHY{location-dependent variation} of the emission rate distribution among different sub-beams of the same beam. We model the $\psi$-dependence as a Power Law with a Super-/Sub-Exponential Cutoff (PLSEC):
\begin{equation}
    C(\psi,\varepsilon)=N(\varepsilon) \left(\frac{\psi}{1\degree}\right)^{-\Gamma(\varepsilon)} \exp{\left[-\left(\frac{\psi}{\Psi_c(\varepsilon)}\right)^{\beta}\right]} \ \ \ .
\end{equation}

We assign the log-prefactor $\ln{N}$ a polynomial of $\ln{\varepsilon}$ with degree 4:
\begin{equation}\label{N_delta}
    \ln{N(\varepsilon)}=\ln{N_0}+\eta_1\ln{\varepsilon}+\eta_2(\ln{\varepsilon})^2+\eta_3(\ln{\varepsilon})^3+\eta_4(\ln{\varepsilon})^4 \ \ \ ,
\end{equation}
the power-law index $\Gamma$ a polynomial of $\ln{\varepsilon}$ with degree 3:
\begin{equation}
    \Gamma(\varepsilon)=\Gamma_0+\Gamma_1\ln{\varepsilon}+\Gamma_2(\ln{\varepsilon})^2+\Gamma_3(\ln{\varepsilon})^3 \ \ \ ,
\end{equation}
the logarithm of the cutoff angle $\Psi_c$ (measured from the beam axis) a parabolic function of $\ln{\varepsilon}$:
\begin{equation}\label{Psi_c_delta}
    \ln{\Psi_c(\varepsilon)}=\ln{\Psi_{c0}}+\alpha_1\ln{\varepsilon}+\alpha_2(\ln{\varepsilon})^2 \ \ \ ,
\end{equation}
and the sharpness index $\beta$ (larger values mean sharper cutoffs) a  "locally uniform"\footnote{\label{LocUni} "Locally uniform" means that a distribution is assumed to be energy-independent for an individual phaseogram, but could be different among different phaseograms of the same pulsar.} distribution.
It is worth mentioning that the power of 3 in Eqn.~\ref{DSint}, associated with the Doppler effect on the “specific intensity”, has been incorporated into the parameter $\eta_1$ here, in principle (detailed explanations are presented in Appendix~\ref{ApxIntensity}).

   \begin{figure}
   \centering
   \includegraphics[width=.99\columnwidth]{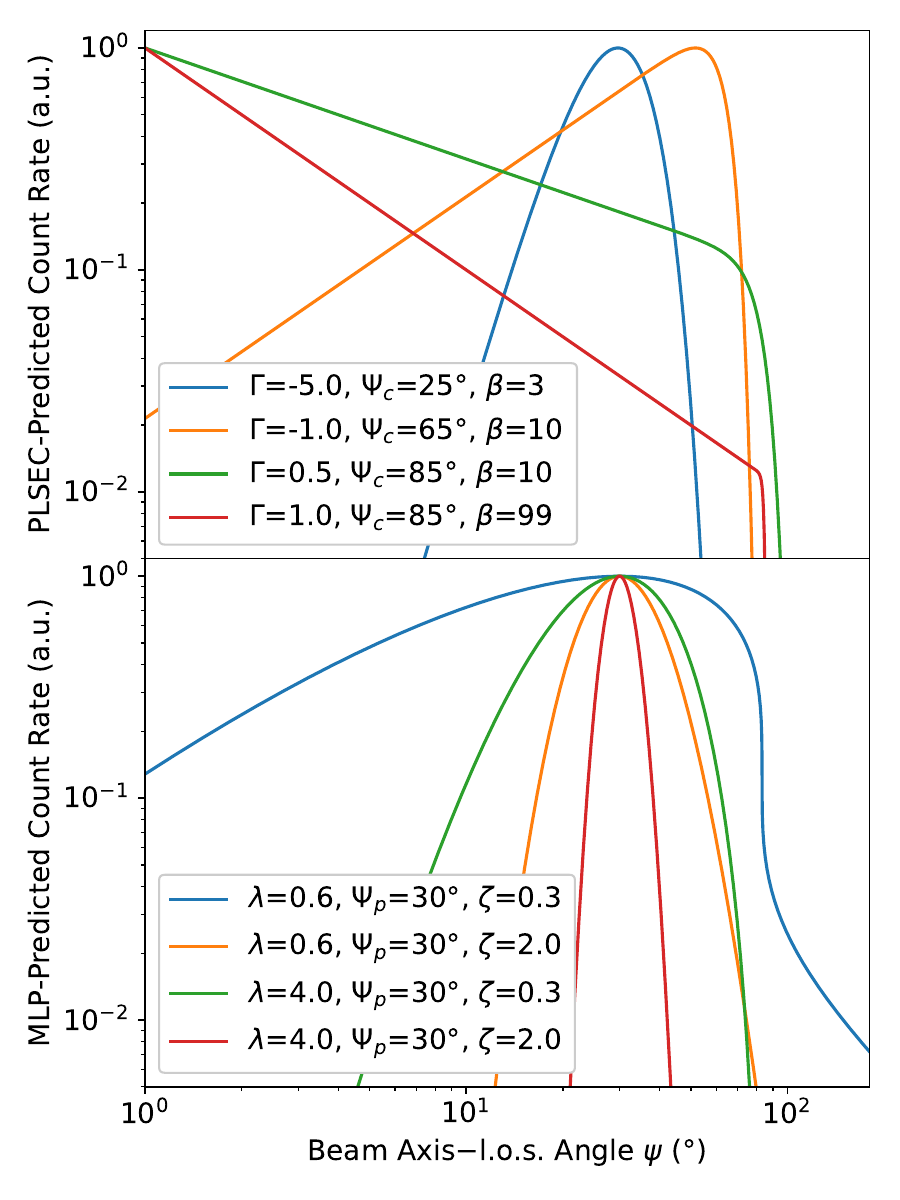}
   \caption{Top panel: Examples of count rate distributions simulated with PLSEC functions. Blue and orange curves demonstrate hollow (i.e. annular) conic beams, while green and red curves demonstrate centrally peaked beam. Bottom panel: Examples of MLP-simulated count rate distributions, with hollow conic structures of the same $\Psi_p$.
}
              \label{BeamShape_examples}%
    \end{figure}

$N$ is, by definition, the count rate extrapolated at $\psi=1\degree$ that is generally unobservable. $\Gamma$ is the power-law index for an inner region of a beam, where we also have limited amount of information. After all, our toy model infers the energy-dependent beam shape based on only an outskirt of the beam. A positive value of $\Gamma$ implies a centrally peaked beam, whose count rate diverges to infinity as $\psi$ approaches 0. A negative value of $\Gamma$, on the contrary, predicts that the count rate declines to 0 as $\psi$ approaches 0, forming a hollow conic structure of the beam. \PKHY{The top panel of Figure~\ref{BeamShape_examples} shows some examples of beam shapes simulated with PLSEC.}

We crosscheck the fittings with a modified-log-parabola (MLP) $\psi$-dependence\footnote{$sgn(x)=1$ for $x\ge0$, while $sgn(x)=-1$ for $x<0$.}:
\begin{equation}
    C(\psi,\varepsilon)=N(\varepsilon) \left(\frac{\psi}{1\degree}\right)^{\lambda(\varepsilon)\mathrm{sgn}{\left(\ln{\frac{\Psi_p^{\zeta+1}(\varepsilon)}{\psi}}\right)}\left|\ln{\frac{\Psi_p^{\zeta+1}(\varepsilon)}{\psi}}\right|^{\zeta}} \ \ \ .
\end{equation}
This function always simulates an annular conic beam. Here, we have performed a transformation of parameters such that $\Psi_p$ is the angular radius with the maximum count rate, while both $\lambda$ and $\zeta$ govern the annular conic thickness and sharpness (larger values imply a thinner and sharper wall of a cone). \PKHY{Some examples of beam shapes simulated with MLP are demonstrated in the bottom panel of Figure~\ref{BeamShape_examples}.}

$N$, the count rate extrapolated at $\psi=1\degree$, is assigned with the same $\varepsilon$-dependence as in Eqn.~\ref{N_delta}. $\lambda$ is assigned with a polynomial of $\ln{\varepsilon}$ with degree 3:
\begin{equation}\label{lambda_delta}
    \lambda(\varepsilon)=\lambda_0+\lambda_1\ln{\varepsilon}+\lambda_2(\ln{\varepsilon})^2+\lambda_3(\ln{\varepsilon})^3 \ \ \ ,
\end{equation}
the logarithm of $\Psi_p$ is assigned with a parabola of $\ln{\varepsilon}$:
\begin{equation}\label{Psi_p_delta}
    \ln{\Psi_p(\varepsilon)}=\ln{\Psi_{p0}}+\alpha_1\ln{\varepsilon}+\alpha_2(\ln{\varepsilon})^2 \ \ \ ,
\end{equation}
and  $\zeta$ is assigned with a  "locally uniform"$^{\ref{LocUni}}$ distribution. With these definitions, PLSEC $\psi$-dependence and MLP $\psi$-dependence contribute an equal number of d.o.f. to the model.

Regardless of the $\psi$-dependent function, the last parameter of this model is the overall background count rate ($C_{bkg}$; i.e. the un-pulsed count rate) of a phaseogram. Adding up the $C$ of beams and $C_{bkg}$ yields the total predicted count rate as a function of phase.

\subsection{Procedures of modeling a phaseogram}

Our toy model is dedicated to fitting its predicted count distribution to observed phaseograms of pulsars. The first step of our algorithm is to convert $t_{det}$ to $t_{psr}$, according to time delay (Eqn.~\ref{TDeqn}). Based on emission geometry, we compute $\psi$ as a function of $t_{psr}$ (Eqn.~\ref{GEOeqn}). Considering Doppler shift, we compute $\varepsilon$ as a function of $t_{psr}$ (Eqn.~\ref{DSeqn}). By substituting $\psi$ and $\varepsilon$, we simulate the count rate as a function of $t_{det}$ and compare this with the detected counts of a phaseogram. We iterate the parameter values until the maximum-likelihood combination is determined.

The geometrical parameters and their definitions are tabulated in Table~\ref{ParDef}. For the count rate distribution of each pulsar, we attempt both PLSEC $\psi$-dependence and MLP $\psi$-dependence, and we focus on the more preferable one.  
In order to reduce the degeneracies in the fittings and the correlations among parameters, for each pulsar, we fit three phaseograms of different energy bands simultaneously, and unify its $\Theta_A$  that we deem to be energy-independent. In this way, the fitting results for three phaseograms of the same pulsar are forced to share the same $\Theta_A$ value, while each of the 21 other parameters may have different values among different phaseograms.

\section{Datasets for case studies}

   \begin{figure*}
   \centering
   \includegraphics[width=.33\textwidth]{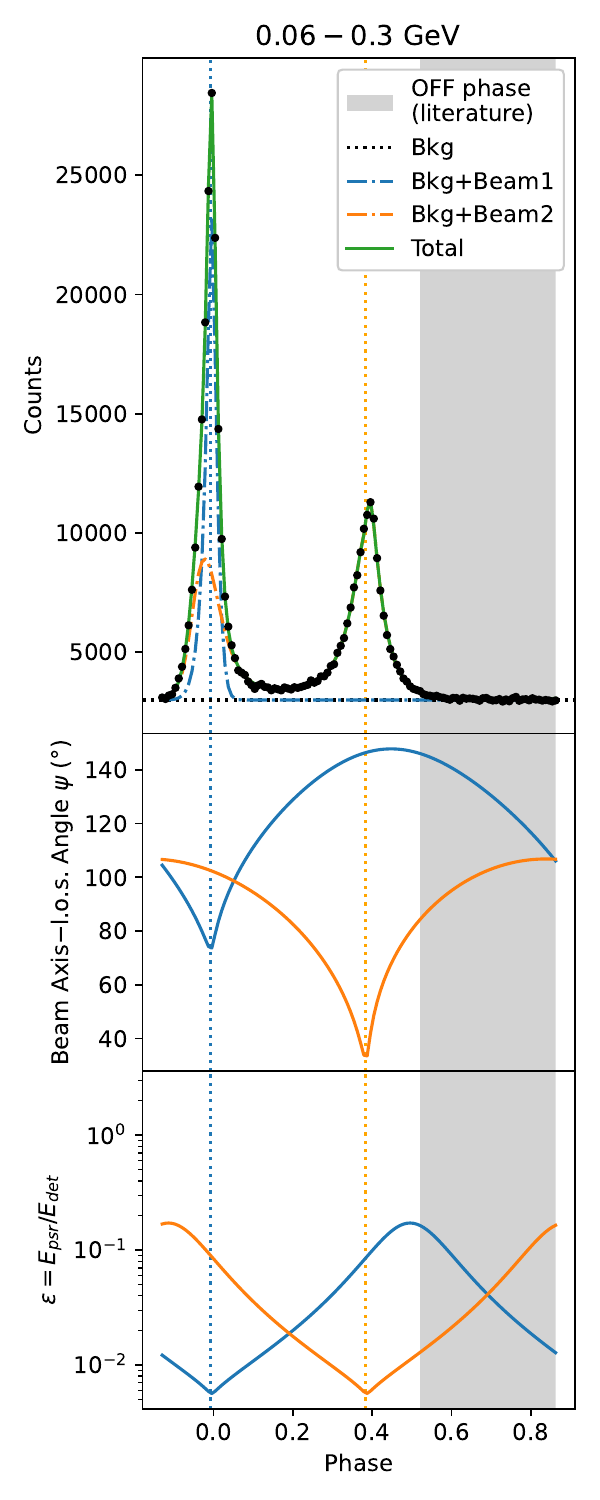}
   \includegraphics[width=.33\textwidth]{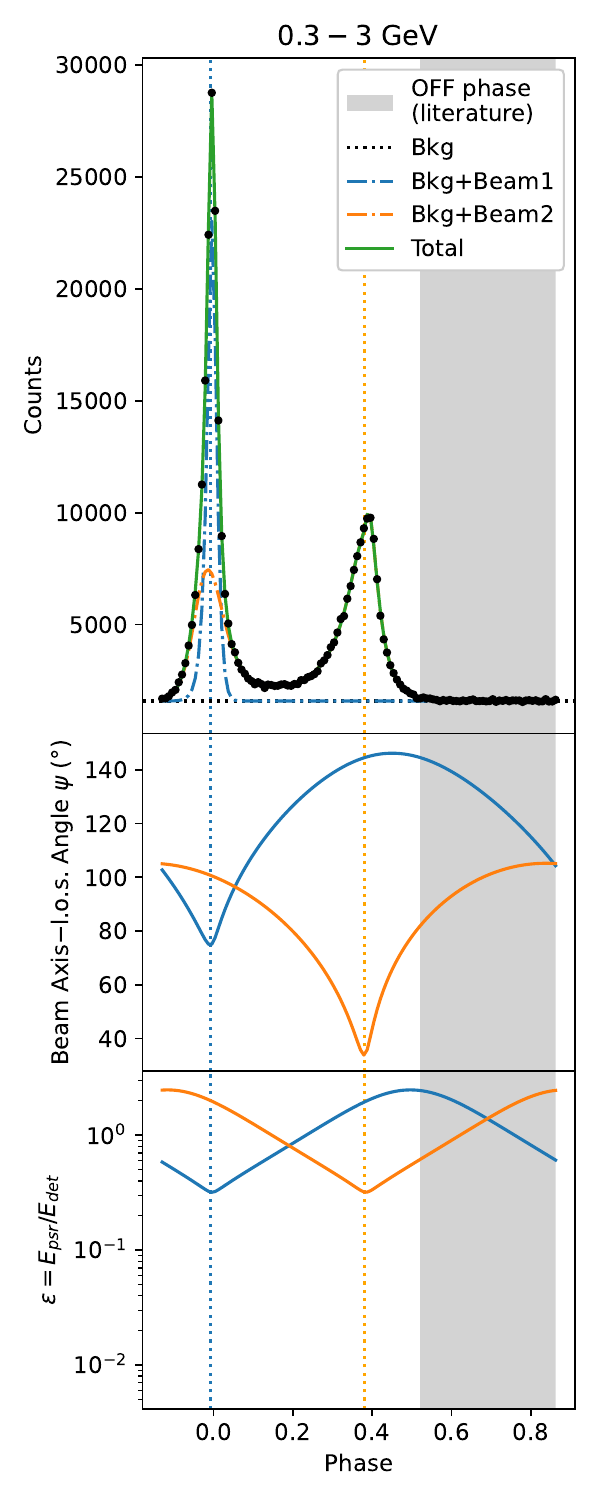}
   \includegraphics[width=.33\textwidth]{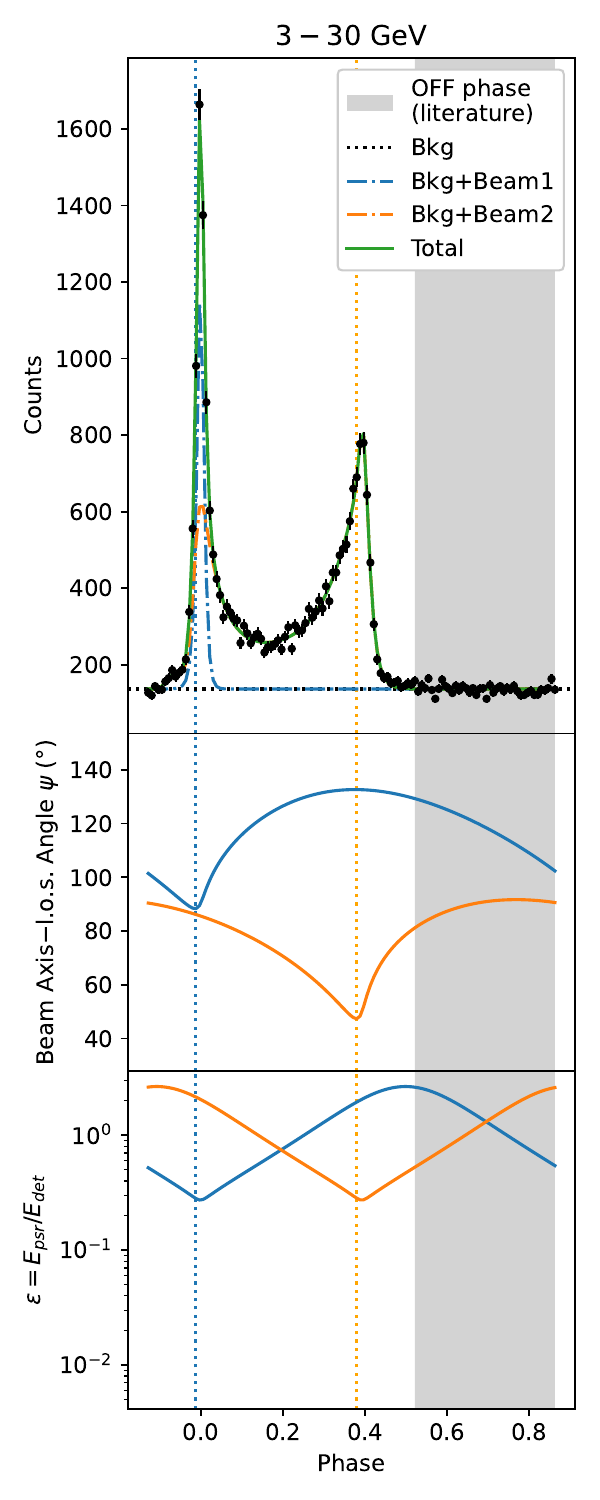}
   \caption{Fitting results of the Crab pulsar, with the more preferable model of PLSEC $\psi$-dependence. Top panels: \emph{Fermi}-LAT $\gamma$-ray phaseograms over specific energy ranges. Middle panels: Model-predicted angle between a beam axis and our l.o.s. ($\psi$), as a function of pulse phase. Bottom panels: Model-predicted inverse Doppler factor ($\varepsilon=E_{psr}/E_{det}$), as a function of pulse phase. Each vertical dotted line spanning over a column of panels indicates the pulse phase when a beam axis is closest to our l.o.s. ($t_1$ or $t_2$; not to be confused with the actually observed peak phase, which is further modulated by the Doppler shift). For all panels, the blue and orange colours refer to Beam1 and Beam2 respectively. The grey shaded interval [0.52, 0.87] was defined as the off-pulse phase range in previous literature \citep{Aleksic_Gap_2012}. 
}
              \label{CrabLC}%
    \end{figure*}

   \begin{figure*}
   \centering
   \includegraphics[width=.33\textwidth]{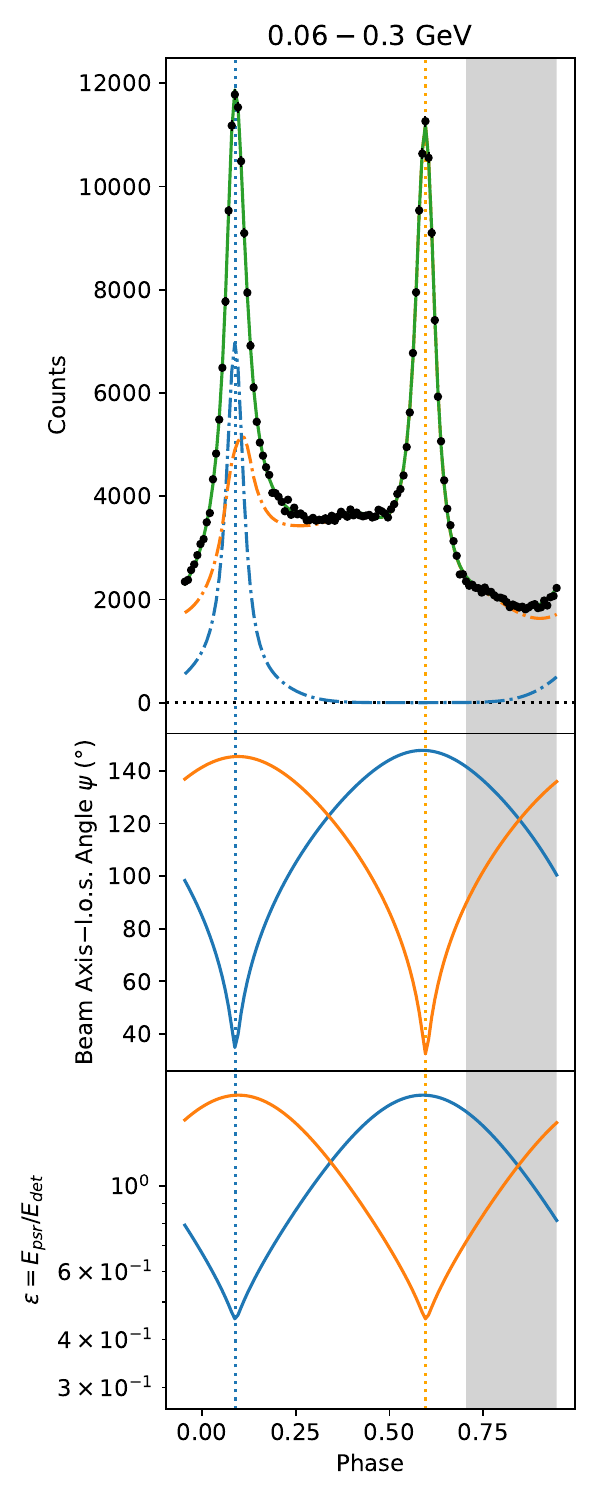}
   \includegraphics[width=.33\textwidth]{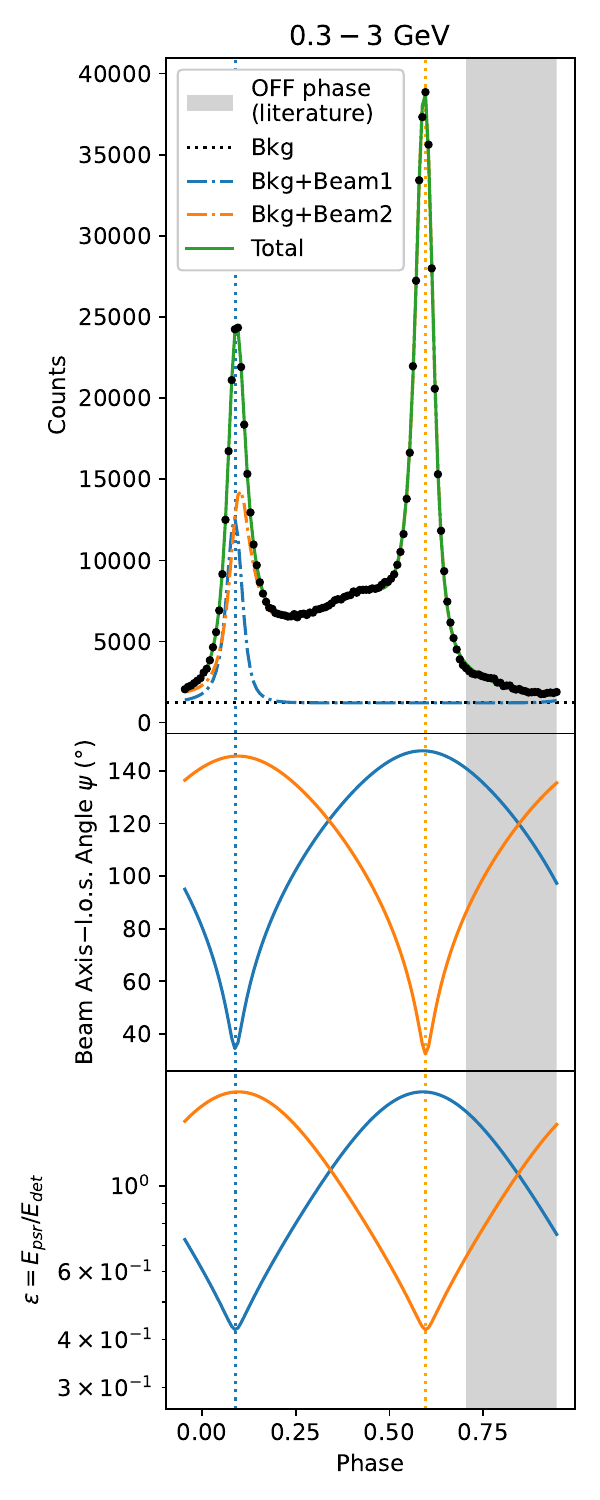}
   \includegraphics[width=.33\textwidth]{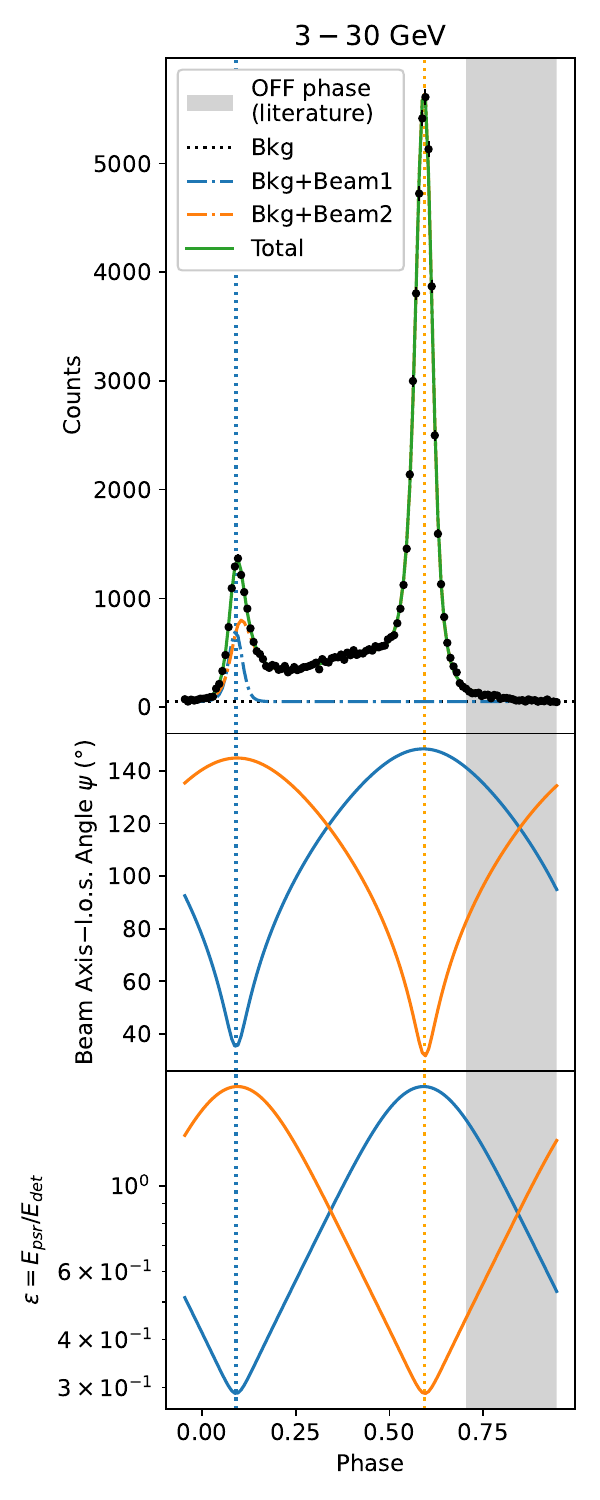}
   \caption{Fitting results of the Geminga pulsar, with the more preferable model of MLP $\psi$-dependence. The panels, colours and line styles have the same meanings as in Fig.~\ref{CrabLC}. \citet{MAGIC_GemingaPSR_2020} defined the grey shaded interval [0.70, 0.95] as the so-called off-pulse phase range. 
}
              \label{GemingaLC}%
    \end{figure*}

We perform case studies on Crab, Geminga, Dragonfly and Vela pulsars, by fitting our prototypical toy model to their \emph{Fermi}-LAT $\gamma$-ray phaseograms. We reconstruct the 0.06$-$0.3~GeV, 0.3$-$3~GeV and 3$-$30~GeV phaseograms of Crab and Geminga pulsars (top panels of Fig.~\ref{CrabLC}~\&~\ref{GemingaLC}), with a circular extraction aperture of $3\degree$ radius centered on the targeted pulsar. For these two pulsars, we adopt the data accumulated over the first 15.7~yr of \emph{Fermi}-LAT observations. The pulsar’s rotational phase corresponding to each detected photon is computed by the PINT Python package \citep[v0.9.7; ][]{Luo2019, Luo2021}, referring to appropriate ephemerides\footnote{Ephemeris of the Crab pulsar provided by Jodrell Bank: \url{https://www.jb.man.ac.uk/pulsar/crab.html}; Ephemeris of the Geminga pulsar provided by G. Ceribella: \url{https://www.mpp.mpg.de/~ceribell/geminga/index.php}}. 

   \begin{figure*}
   \centering
   \includegraphics[width=.33\textwidth]{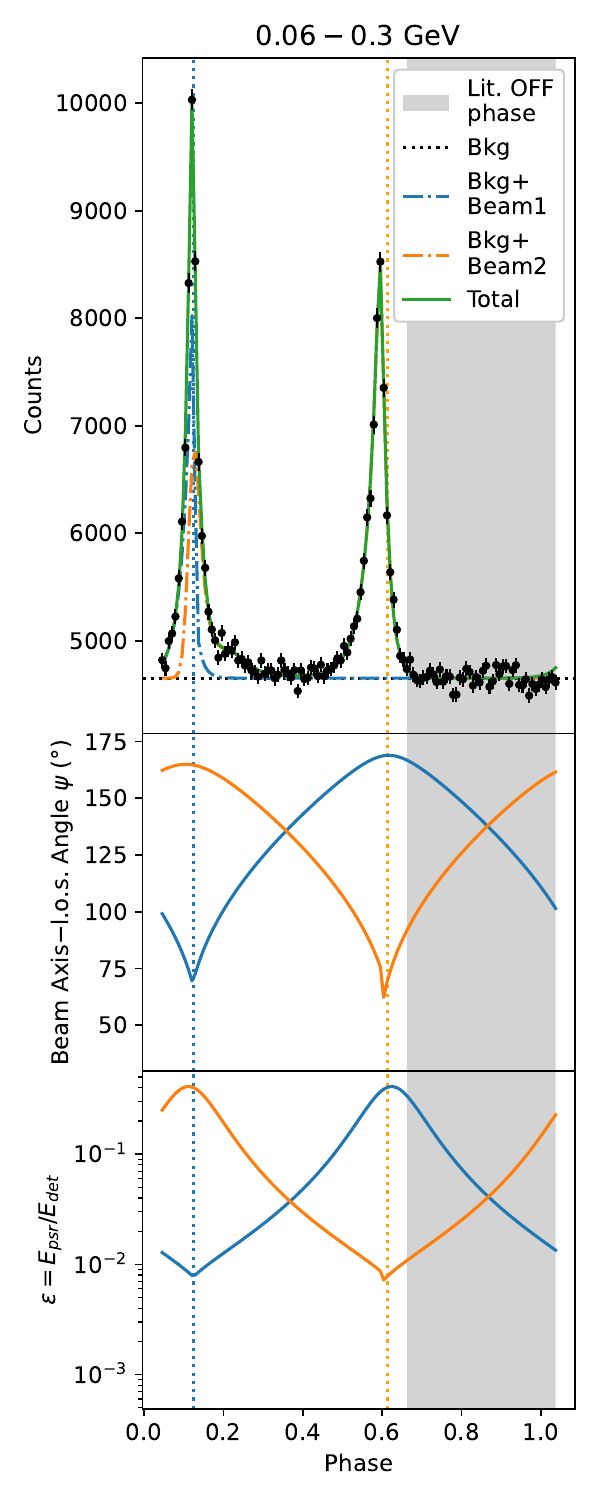}
   \includegraphics[width=.33\textwidth]{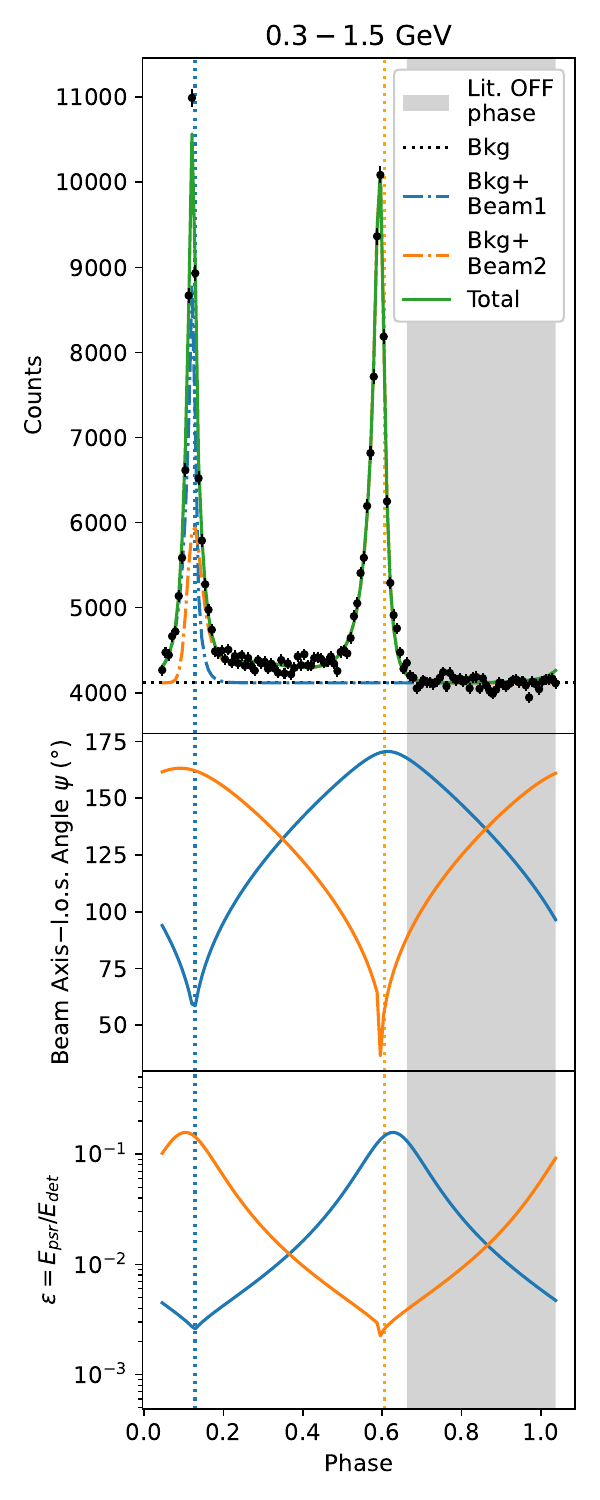}
   \includegraphics[width=.33\textwidth]{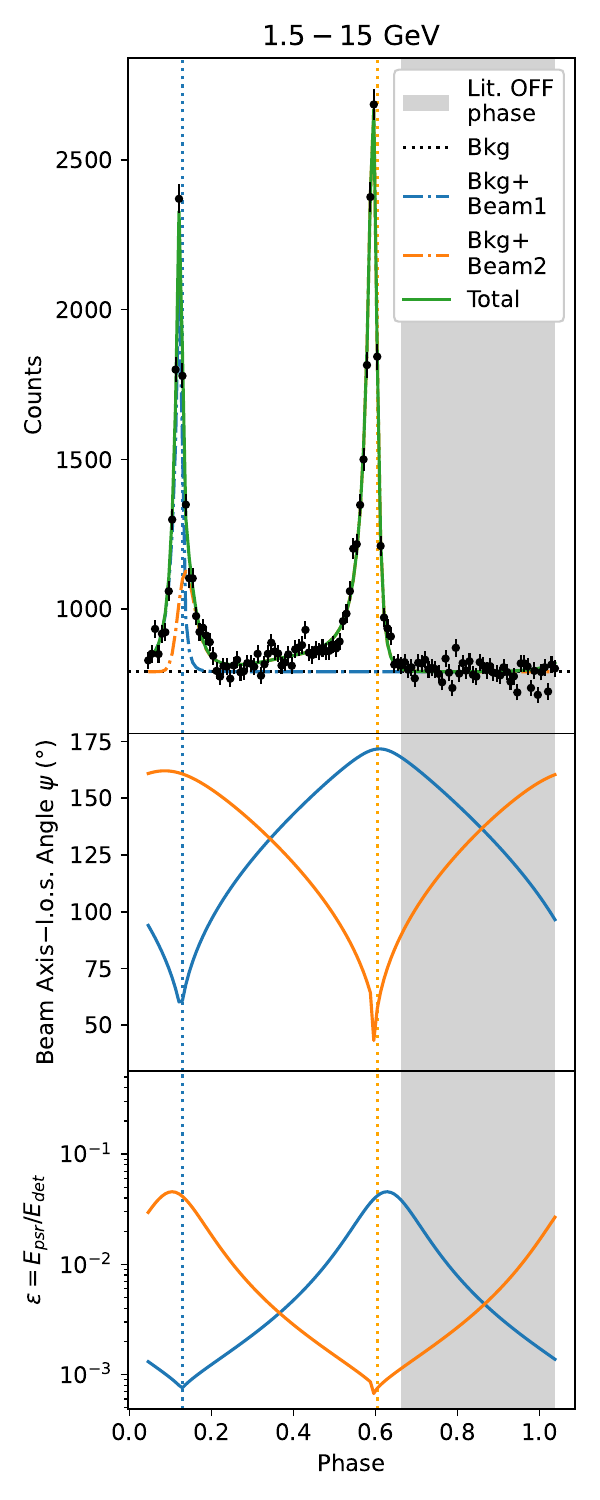}
   \caption{Fitting results of the Dragonfly pulsar, with the more preferable model of MLP $\psi$-dependence. The panels, colours and line styles have the same meanings as in Fig.~\ref{CrabLC}. The grey shaded interval [0.660, 1.045] was defined as the off-pulse phase range in \citet{Wang_thesis_2025}. 
}
              \label{DragonflyLC}%
    \end{figure*}

\PKHY{For the Dragonfly pulsar, we reconstruct its phaseograms (top panels of Fig.~\ref{DragonflyLC}) from a data product attached to the LAT Third Catalog of Gamma-ray Pulsars\footnote{LAT photon data file of the Dragonfly pulsar: \url{https://heasarc.gsfc.nasa.gov/FTP/fermi/data/lat/catalogs/3PC/photon/3deg_50MeV/J2021+3651_3p0deg_50MeV_300000MeV_105deg_128_3.fits}}. We choose these three energy bands: 0.06$-$0.3~GeV, 0.3$-$1.5~GeV and 1.5$-$15~GeV.}

   \begin{figure*}
   \centering
   \includegraphics[width=.33\textwidth]{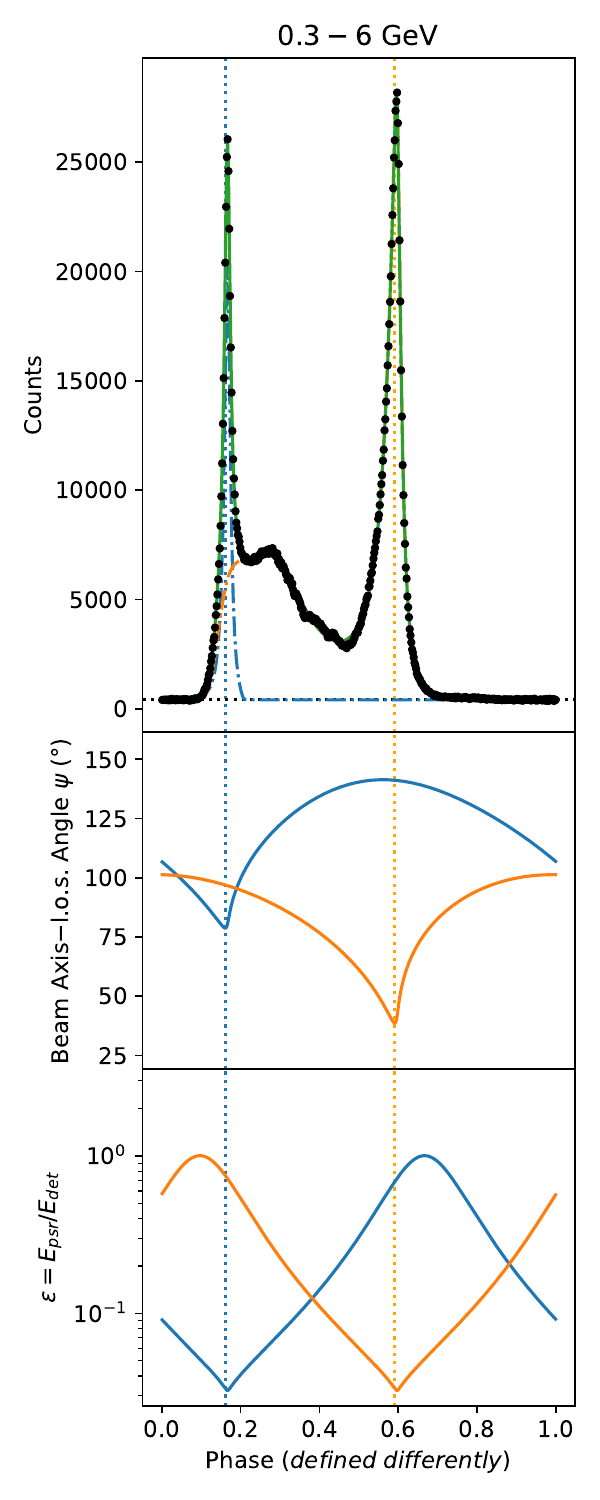}
   \includegraphics[width=.33\textwidth]{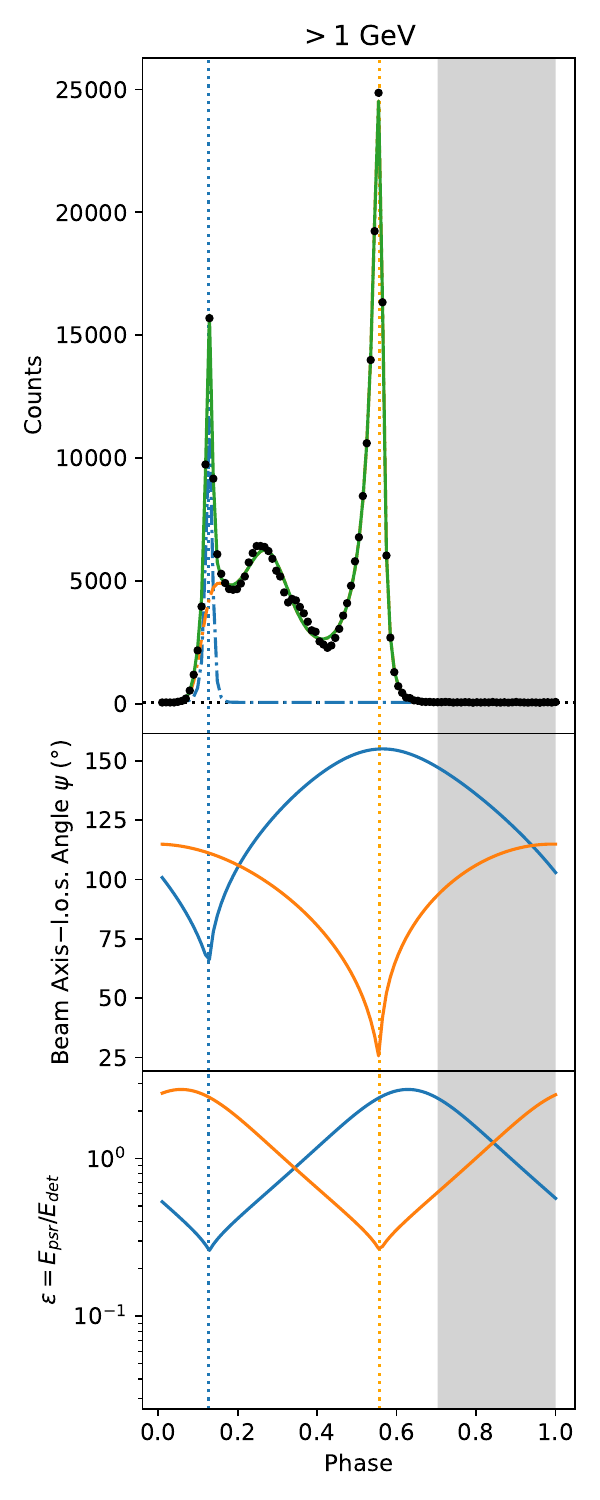}
   \includegraphics[width=.33\textwidth]{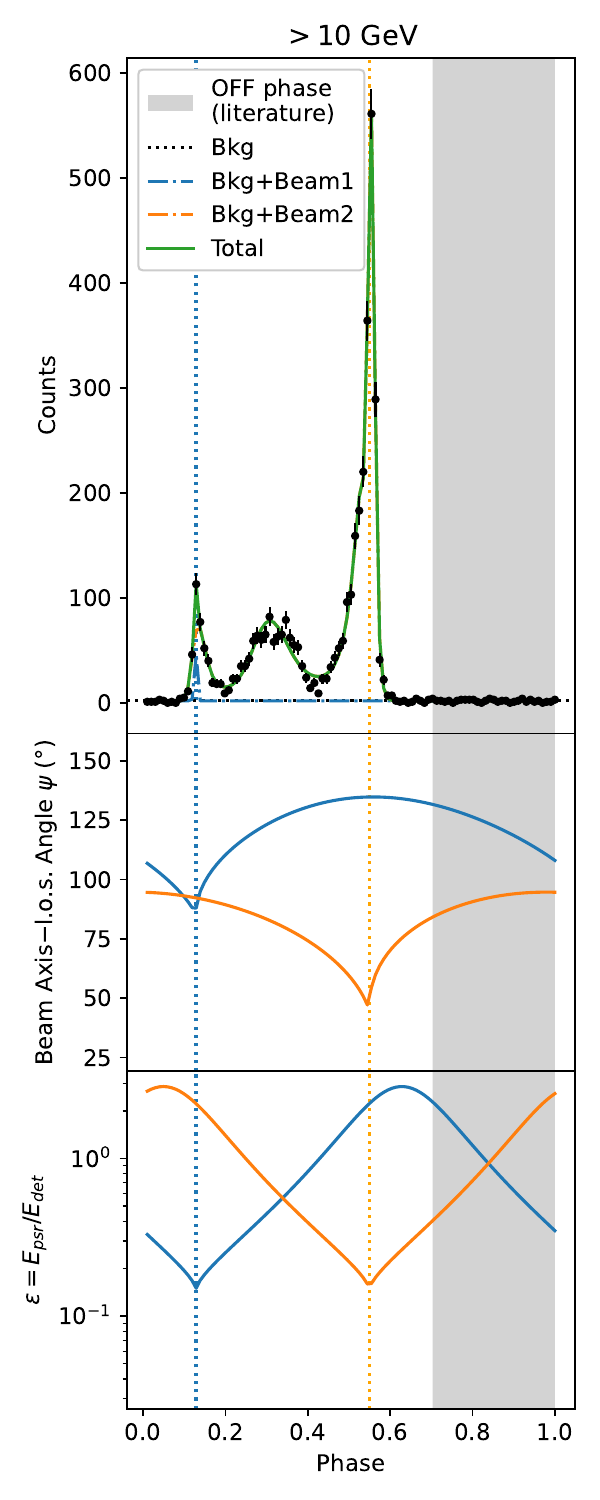}
   \caption{Fitting results of the Vela pulsar, with the more preferable model of PLSEC $\psi$-dependence. The panels, colours and line styles have the same meanings as in Fig.~\ref{CrabLC}. The data of the 0.3$-$6~GeV phaseogram are taken from \citet{Kargaltsev2023}, while the data of the $>1~$GeV and $>10~$GeV phaseograms are taken from \citet{HESS_Vela_2023}. The off-pulse phase range [0.7, 1.0] adopted by \citet{HESS_Vela_2023} was originally defined in \citet{HESS_Vela_2018}. The 0.3$-$6~GeV phaseogram is reconstructed with a different ephemeris having a different definition of the phase 0.
}
              \label{VelaLC}%
    \end{figure*}

For the Vela pulsar, we work on the 0.3$-$6~GeV phaseogram published in \citet{Kargaltsev2023} in addition to the $>1~$GeV and $>10~$GeV phaseograms published in \citet{HESS_Vela_2023} (top panels of Fig.~\ref{VelaLC}). These three energy bands overlap with one another, leading to double counts of some photons in the whole sample. Nevertheless, we deem the caused artificial effect to be negligible, because those double counts only lead to further underestimation of the already tiny statistical error of $\Theta_A$. We note that the 0.3$-$6~GeV phaseogram is reconstructed with a different ephemeris having a different definition of the phase 0, resulting that its phase values are larger by 0.035--0.045 with respect to the two other phaseograms.  

\REFE{The selections of all aforementioned data samples are regardless of photon weights -- the probability that each photon is associated with a particular source. It is actually unnecessary to apply a photon-weight cut in the data screening procedure for this work, because the un-pulsed term $C_{bkg}$ of our toy model is dedicated to accounting for photons unassociated with the targeted pulsar.}

\section{Results and discussion of case studies}

\begin{table*}
	\centering
	\caption{Parameter values of the more preferable models for Crab and Vela pulsars, where their count rate distributions are modelled with PLSEC $\psi$-dependence. }
	\label{PLSEC_values}
        
\begin{tabular}{cc|ccc|ccc}
 \hline      &          &        \multicolumn{3}{c}{Crab}                              & \multicolumn{3}{c}{Vela}                      \\
 \hline  $E_{det}$                             & (GeV)    & 0.06$-$0.3 & 0.3$-$3  & 3$-$30   & 0.3$-$6       & $>$1 & $>$10 \\
 \hline \multicolumn{8}{c}{\underline{Iterated geometrical parameters}}                  \\
$\Theta_A$                                & ($\degree$)    & \multicolumn{3}{c}{69.5}     & \multicolumn{3}{c}{70.0}   \\
$d\sin{\theta_B}$                         & ($R_{LC}$)    & 0.96     & 0.88   & 0.89   & 0.98        & 1.02            & 1.07             \\
$d\cos{\theta_B}$                         & ($R_{LC}$)    & 0.98     & 1.01   & 0.96   & 0.64        & 0.65            & 0.73             \\
$t_0$                                & (phase)  & 0.102    & 0.115  & 0.115  & 0.226$-$0.236* & 0.226           & 0.219            \\
$\theta_C$                                & ($\degree$)    & 37.4     & 35.7   & 22.2   & 31.3        & 45.0            & 24.8             \\
$\theta_Q$                                & ($\degree$)    & 99.1     & 99.1   & 114.5  & 110.0       & 101.8           & 102.6            \\
$\gamma_{Lor}$               &               & 91.6     & 1.8      & 2.0    & 15.9   & 2.1         & 3.6                              \\
$\sigma_{ver}$                              & ($R_{LC}$)    & 0.153    & 0.014  & 0.004  & 0.114       & 0.204           & 0.194            \\
\multicolumn{8}{c}{\underline{Additional geometrical properties}}                  \\
$d$                                 & ($R_{LC}$)    & 1.37     & 1.34   & 1.31   & 1.17        & 1.21            & 1.29             \\
${\theta_B}$                                & ($\degree$)    & 44.6     & 41.2   & 43.1   & 56.7        & 57.6            & 55.8             \\
$t_1$                                & (phase)  & $-$0.007   & $-$0.008 & $-$0.015 & 0.116$-$0.126* & 0.127           & 0.128            \\
$t_2$                                & (phase)  & 0.384    & 0.380  & 0.379  & 0.546$-$0.556* & 0.556           & 0.549            \\
$\sigma_{rad}$                              & ($R_{LC}$)    & 0.04     & $-$0.05$^{\dag}$  & $-$0.02$^{\dag}$  & 0.02        & ---$^{\ddag}$              & ---$^{\ddag}$               \\
\multicolumn{8}{c}{\underline{Energy-dependent beam shape}}                  \\
$N_0$$^{\diamond}$                                & (a.u.)   & ---       & ---     & ---     & ---          & ---              & ---               \\
$\eta_1$                &                & 97.1     & 185.9    & 248.7  & 170.5  & 305.9       & $-$75.4                            \\
$\eta_2$                &                & 8.0      & 34.0     & 46.5   & 42.1   & 26.1        & $-$119.4                           \\
$\eta_3$                 &               & 0.06     & $-$10.81   & 0.46   & 8.45   & $-$34.04      & 48.74                            \\
$\eta_4$                 &               & 0.010    & 2.581    & 1.368  & 0.799  & 0.565       & 15.324                           \\
$\Gamma_0$                &                & 46.4     & 32.4     & 55.5   & 69.8   & 61.4        & 61.9                             \\
$\Gamma_1$                &                & 13.4     & 39.1     & 54.7   & 34.6   & 60.0        & $-$54.4                            \\
$\Gamma_2$                 &               & 0.79     & 4.61     & 8.84   & 6.91   & $-$0.61       & $-$57.37                           \\
$\Gamma_3$                &                & $-$0.004   & $-$3.271   & 0.075  & 0.790  & $-$8.546      & $-$3.089                           \\
$\Psi_{c0}$                               & ($\degree$)    & 67.5     & 94.4   & 134.1  & 96.9        & 81.8            & 65.5             \\
$\alpha_1$                &                & $-$0.126   & $-$0.026   & $-$1.117 & 0.007  & $-$0.131      & $-$0.056                           \\
$\alpha_2$                 &               & $-$0.015   & $-$0.076   & 0.709  & 0.007  & 0.059       & 0.005                            \\
$\beta$                   &              & 11.32    & 13.98    & 152.00 & 77.38  & 9.94        & 15.10                            \\
\multicolumn{8}{c}{\underline{Overall background}}                  \\
$C_{bkg}$                              & (counts) & 2990.0   & 1589.0 & 136.6  & 411.1       & 62.6            & 1.9             \\  \hline 
\end{tabular}

* Corrected for discrepant definitions of phase 0 between the two ephemerides.  \\
$^{\dag}$ The emission site has its most representative location within the light cylinder, 	 but extends beyond the light cylinder, resulting in a negative $\sigma_{rad}$.  \\
$^{\ddag}$ The most representative location of the emission site is outside the light cylinder, so that $\sigma_{rad}$ does not reflect the genuine radial extension at all.  \\
$^{\diamond}$ The count rate is related to not only the pulsar’s properties, but also instrumental properties, data selection criteria and the pulsar’s distance from us. In view of such an ambiguous physical meaning, we skip reporting its normalisation factor.
\end{table*}

\begin{table*}
	\centering
	\caption{Parameter values of the more preferable model for the Geminga and Dragonfly pulsars, where their count rate distributions are modelled with MLP $\psi$-dependence. }
	\label{MLP_values}
        
\begin{tabular}{cc|ccc|ccc}
 \hline      &          &        \multicolumn{3}{c}{Geminga}           &        \multicolumn{3}{c}{Dragonfly}           \\
 \hline  $E_{det}$                              & (GeV)    & 0.06$-$0.3 & 0.3$-$3  & 3$-$30  & 0.06$-$0.3 & 0.3$-$1.5 & 1.5$-$15    \\
 \hline \multicolumn{8}{c}{\underline{Iterated geometrical parameters}}                  \\
$\Theta_A$                                & ($\degree$)    & \multicolumn{3}{c}{56.7}        & \multicolumn{3}{c}{77.0}      \\
$d\sin{\theta_B}$                         & ($R_{LC}$)    & 1.08     & 0.99   & 0.90  & 1.31                             & 1.19                            & 1.20                           \\
$d\cos{\theta_B}$                         & ($R_{LC}$)    & $-$0.05    & $-$0.04  & $-$0.02 & 0.17                             & 0.33                            & 0.33                           \\
$t_0$                                & (phase)  & 0.196    & 0.208  & 0.222 & 0.172                            & 0.194                           & 0.192                          \\
$\theta_C$                                & ($\degree$)    & 88.8     & 89.0   & 88.3  & 88.0                             & 86.3                            & 85.2                           \\
$\theta_Q$                                & ($\degree$)    & 90.2     & 90.2   & 90.2  & 91.1                             & 92.2                            & 93.0                           \\
$\gamma_{Lor}$               &               & 1.4      & 1.5      & 2.0      & 94.2                             & 245.8                           & 850.0                              \\
$\sigma_{ver}$                              & ($R_{LC}$)    & 0.19     & 0.18   & 0.18  & 0.22                             & 0.23                            & 0.26                           \\
\multicolumn{8}{c}{\underline{Additional geometrical properties}}                  \\
$d$                                 & ($R_{LC}$)    & 1.08     & 0.99   & 0.90  & 1.32                             & 1.23                            & 1.24                           \\
${\theta_B}$                                & ($\degree$)    & 92.5     & 92.5   & 91.3  & 82.6                             & 74.6                            & 74.5                           \\
$t_1$                                & (phase)  & 0.089    & 0.089  & 0.091 & 0.125                            & 0.129                           & 0.129                          \\
$t_2$                                & (phase)  & 0.597    & 0.597  & 0.594 & 0.613                            & 0.605                           & 0.605                          \\
$\sigma_{rad}$                              & ($R_{LC}$)    & ---$^{\ddag}$       & $-$0.26$^{\dag}$  & $-$0.03$^{\dag}$  & ---$^{\ddag}$  & ---$^{\ddag}$  & ---$^{\ddag}$ \\
\multicolumn{8}{c}{\underline{Energy-dependent beam shape}}                  \\
$N_0$$^{\diamond}$                                & (a.u.)   & ---       & ---     & ---  & ---       & ---     & ---      \\
$\eta_1$                &                & 51.1     & 95.5     & 122.4     & 89.4                             & 60.0                            & 46.0   \\
$\eta_2$                &                & 107.4    & 221.8    & 160.1     & $-$12.4                            & $-$8.8                            & $-$6.3   \\
$\eta_3$                 &               & $-$30.89   & 1.05     & 0.28     & 0.80                             & 0.31                            & 0.25    \\
$\eta_4$                 &               & $-$14.240  & $-$1.365   & $-$0.075    & 0.457                            & 0.205                           & 0.107   \\
$\lambda_0$                &                & 53.7     & 139.1    & 286.9     & 41.7                             & 48.3                            & 54.2   \\
$\lambda_1$                 &               & 96.6     & 226.0    & 365.7     & $-$15.1                            & 15.3                            & 11.9   \\
$\lambda_2$                 &               & 43.77    & 86.06    & 113.03    & $-$8.83                            & 1.74                            & 0.46   \\
$\lambda_3$                 &               & 3.621    & 0.185    & $-$0.501    & $-$0.879                           & 0.077                           & $-$0.027   \\
$\Psi_{p0}$                               & ($\degree$)    & 11.0     & 11.0   & 10.9  & 2.8                              & 3.0                             & 2.6                            \\
$\alpha_1$                &                & $-$0.024   & 0.014    & 0.044    & $-$0.038                           & 0.051                           & 0.014    \\
$\alpha_2$                 &               & 0.085    & 0.023    & $-$0.005    & $-$0.010                           & 0.004                           & $-$0.001   \\
$\zeta$                   &              & 1.06     & 1.06     & 1.06    & 2.56                             & 2.51                            & 3.14     \\
\multicolumn{8}{c}{\underline{Overall background}}                  \\
$C_{bkg}$                              & (counts) & 0.0      & 1216.0 & 51.1 & 4653.0                           & 4116.0                          & 791.2                         \\  \hline 
\end{tabular}

$^{\dag}$ The emission site has its most representative location within the light cylinder, 	 but extends beyond the light cylinder, resulting in a negative $\sigma_{rad}$.  \\
$^{\ddag}$ The most representative location of the emission site is outside the light cylinder, so that $\sigma_{rad}$ does not reflect the genuine radial extension at all.  \\
$^{\diamond}$ The count rate is related to not only the pulsar’s properties, but also instrumental properties, data selection criteria and the pulsar’s distance from us. In view of such an ambiguous physical meaning, we skip reporting its normalisation factor.
\end{table*}

For Crab and Vela pulsars, a joint likelihood ratio test for a simultaneous fit of three phaseograms indicates that PLSEC $\psi$-dependence of a count rate distribution is much more preferable than MLP $\psi$-dependence (2$\Delta$ln(likelihood) $>4\times10^3$ between two models, with the same number of d.o.f.). Accordingly, we only report the models of PLSEC $\psi$-dependence for Crab and Vela pulsars (Fig.~\ref{CrabLC}~\&~\ref{VelaLC} and Table~\ref{PLSEC_values}). It is interesting to note that our fittings imply abrupt super-exponential cutoffs ($\beta\gtrsim10$) for Crab and Vela pulsars.

\PKHY{For Geminga and Dragonfly pulsars, MLP $\psi$-dependence is preferred over PLSEC $\psi$-dependence with 2$\Delta$ln(likelihood) of $\sim770$ and $\sim330$ respectively, while PLSEC $\psi$-dependence still gives satisfactory fits. The results obtained with MLP $\psi$-dependence are presented in Fig.~\ref{GemingaLC}~\&~\ref{DragonflyLC} and Table~\ref{MLP_values}, while those obtained with PLSEC $\psi$-dependence are briefly discussed in Appendix~\ref{ApxGeminga} for crosschecking some features of the Geminga and Dragonfly pulsars.}

The statistical uncertainty of each parameter is generally tiny compared to the corresponding value, thanks to the huge amount of data. To be specific, the percentage errors (statistical) are $\lesssim$0.03\% for $\Theta_A$ and $\lesssim$0.6\% for $d\sin{\theta_B}$, $t_0$, $\theta_C$, $\theta_Q$ \& $\gamma_{Lor}$. The statistical errors of $d\cos{\theta_B}$ and $\sigma_{ver}$ are $\lesssim6\times10^{-3}~R_{LC}$ and $\lesssim0.17~R_{LC}$ respectively. The systematic uncertainties of parameters are beyond the scope of this methodology paper. Therefore, we skip reporting errors in Tables~\ref{PLSEC_values}~\&~\ref{MLP_values}.

\subsection{Features shared in common}

Our prototypical toy model suggests some common features of Crab, Geminga, Dragonfly and Vela pulsars. \PKHY{We discuss their common features in three aspects: Consistencies with wind models, the orientation of the magnetic pole, and an extensive tail of a pulse component.}

\subsubsection{Consistencies with wind models}\label{WindModels}

For each phaseogram of them, the fitting always yields a $d\sin{\theta_B}$ value within \PKHY{0.88--1.31~$R_{LC}$} (i.e. the most representative location of the emission site is very close to or even slightly beyond the light cylinder wall; relevant clarifications are made in Appendix~\ref{ApxBeyondLC}), and a $\theta_Q$ value within $90\degree$--$115\degree$ (close to a straight angle). These make our results compatible with wind models \citep[e.g.,][]{Aharonian_wind_2012,HESS_Vela_2023}. Notably, a nearly straight angle $\theta_Q$ leads to this interesting prediction: A beam is highest blueshifted when its axis is almost closest to our l.o.s. (i.e. the phase of minimum $\psi$ almost corresponds to the phase of minimum $\varepsilon$), as shown in middle and bottom panels of Fig.~\ref{CrabLC}--\ref{VelaLC}.

\REFE{Furthermore, most of our fitting results entail moderately extended emission sites as well as very broad/flat beams, including those with $\Psi_c>70\degree$ and those with hollow conic shapes. Such a configuration, to some degree, resembles the very diffuse emission from a very extended region of a wind current sheet.}

\subsubsection{Speculation on the magnetic pole}

We recall that the magnetic pole's orientation (although not parametrised in our toy model) might be indirectly speculated in accordance with other geometrical properties. Qualitatively speaking, our findings of $d\sin{\theta_B}{\sim}R_{LC}$ and $\theta_Q\sim90\degree$ might require the magnetic pole to be azimuthally coincident with one beam axis and azimuthally opposite to another beam axis. With such a configuration, the anticipated magnetic field lines would be closely parallel to the beam axes, so that the emissions along/around the beam axes could better survive magnetic absorption of photons (i.e. magnetic pair creation) and/or caustic effects \citep[e.g.][]{Dyks_caustic_2003, Bai_caustic_2010, Petri_caustic_2015}.  

\subsubsection{Magic of Doppler boost}

We highlight a "hoax" that could be played by Doppler shift. \REFE{At different pulse phases, we are observing each beam at not only different $\psi$ but also different $\varepsilon$ (i.e. different source-frame energies $E_{psr}$). We recall that the detected count rate ($C$) of each beam is a function of both $\psi$ and $\varepsilon$. For a fixed value of $\varepsilon$, away from the beam axis (i.e. $\psi$ increases), $C$ generally drops. Whereas, for a fixed value of $\psi$, $C$ may increase or decrease with $\varepsilon$, depending on the energy-dependence of the beam shape. Interestingly, the interplay between the  energy-dependent beam shape and the temporal distribution of $\varepsilon$ could lead to an extensive tail and/or even sub-spikes of a detected pulse.}

\REFE{As predicted by our fitting results for the four targeted pulsars, this effect enables Beam2 to account for the P2 and Bridge/P3 emissions (as well as significant fractions of P1 emissions). On the other hand, under our assumptions of north-south symmetry and a circular beam profile, neither an extensive tail nor sub-spikes is favoured for Beam1. This could lead to the non-detection of the hypothetical "Bridge2/P4". In the following, we elucidate their distinct pulse shapes in a qualitative manner, with reference to the middle-right and bottom-right panels of Fig.~\ref{CrabLC} (a representative example).}

\REFE{Beam2 generates a major peak at phase$\sim$0.4 when both $\psi$ and $\varepsilon$ are almost at their minimum. Additionally, it generates a minor peak at phase$\sim$0.0 when both $\psi$ and $\varepsilon$ are approaching their maximum. The main reason for the minor peak of Beam2 is that $C$ at phase$\sim$0.0 increases with $\varepsilon$ rapidly, overcoming its $\psi$-dependent decrement. Differently, Beam1 generates only one peak at phase$\sim$0.0 when both $\psi$ and $\varepsilon$ are almost at their minimum. It is because our l.o.s. crosses Beam1 at an outer part, where $C$ drops with $\psi$ more furiously.} \REFESEC{The $\psi$-dependent steepening of a beam shape is particularly conspicuous for our assumed PLSEC and MLP functions (Figure~\ref{BeamShape_examples}). Considering a sharply truncated beam profile, a north-south symmetric environment and the symmetry-breaking l.o.s. altogether, one can speculate a very high difficulty in simultaneously producing the Doppler-boosted tail/sub-spikes of both Beam1 and Beam2.}

\REFESEC{It must be clarified that the predictions of concrete pulse features by our model are data-driven and, at a certain level, dependent on our assumptions on the geometry. The crosschecking results in Appendix~\ref{ApxGeminga} prove that an inferred pulse shape is moderately influenced by our assumed beam morphological model. Specifically, for each phaseogram of the Geminga pulsar, the models of PLSEC $\psi$-dependence and MLP $\psi$-dependence predict different amplitudes and phases for the minor peak of Beam2. Therefore, some concrete predictions, such as non-detection of hypothetical "Bridge2/P4" and over-domination of Vela pulsar's $>10~$GeV emissions by Beam2, are not confirmed yet and require further crosschecks with an advanced  model of a more complicated geometry (proposed in \S\ref{Asym_beam}--\S\ref{MultipleComponents}).}

\subsection{Crab pulsar}

Our fittings for the 0.06$-$0.3~GeV and 0.3$-$3~GeV phaseograms of the Crab pulsar suggest a roughly energy-independent distribution of $\Psi_c$ ($80\degree$--$95\degree$, as shown in the middle-right panel of Fig.~\ref{Crab_check}). Differently, our result for the 3$-$30~GeV phaseogram predicts that $\Psi_c$ first jumps to $>180\degree$ and then decreases furiously with $\varepsilon$. Coincidently, the phase-resolved spectral energy distributions of the Crab pulsar reveal that its spectral shape transits from a sub-exponential cutoff to a power-law tail at around 10~GeV, hinting at a change of the dominating radiation mechanism from curvature radiation to inverse-Compton scattering \citep[e.g.][]{ansoldi_teraelectronvolt_2016, Yeung_CrabPSR_2020, CTAO-LST_CrabPSR_2024}. We speculate that such a transition is qualitatively associated with the dramatic jump of $\Psi_c$ for the 3$-$30~GeV phaseogram. 

\citet{Ng_Tori_2004} characterised the termination shock (torus) of the Crab’s pulsar wind nebula (PWN) system through imaging. It is comforting to note that they constrained the inclination angle of its polar axis from our l.o.s. to be $61\degree$--$63\degree$, that is in a rough agreement with our determined $\Theta_A=69.5\degree$ for the Crab pulsar.

\subsubsection{Correlation with the IXPE results on X-ray polarisation}

   \begin{figure}
   \centering
   \includegraphics[width=\columnwidth]{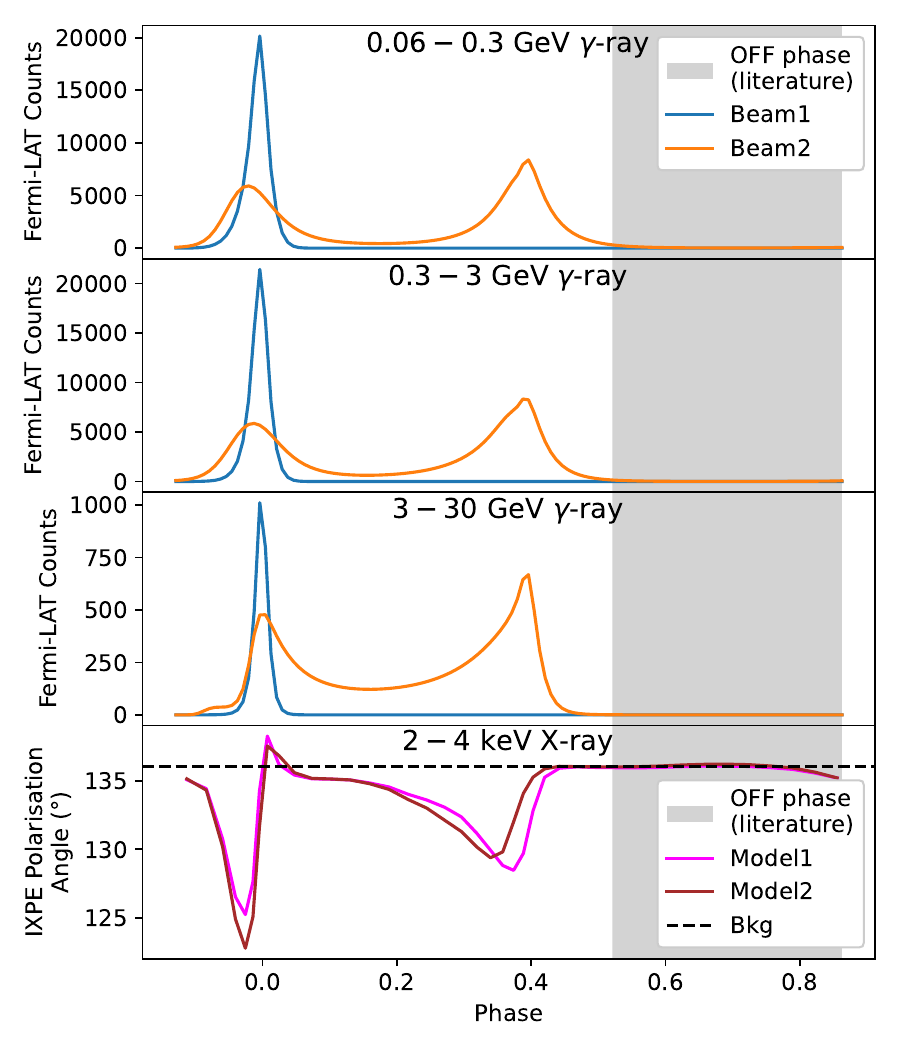}
   \caption{Comparison of our model-predicted $\gamma$-ray pulse shapes with the X-ray polarisation angles, for the Crab pulsar. The model lines of the X-ray polarisation angles are taken from Fig.~2 of \citet{Gonzalez2025}, who fitted the IXPE data under the theoretical framework of \citet{Slowikowska2009}. The grey shaded interval [0.52, 0.87] was defined as the off-pulse phase range in previous literature \citep{Aleksic_Gap_2012}. 
}
              \label{Counts_vs_PolAngles}%
    \end{figure}

For each phaseogram of the Crab pulsar, Beam1’s peak significatly overlaps with Beam2’s Doppler-boosted tail. We make an interesting comparison of our model-predicted $\gamma$-ray pulse shapes with the X-ray polarisation angles (Fig.~\ref{Counts_vs_PolAngles}). \citet{Gonzalez2025} fitted the IXPE data with a model modified from that of \citet{Slowikowska2009}, describing the X-ray polarisation angle as a function of pulse phase.

Regardless of the energy band, the Beam1 predicted by our model only dominates the $\gamma$-ray emission at the TW of P1. Coincidently, the X-ray polarisation angle is larger than the background level only at the TW of P1. The X-ray polarisation angle is smaller than the background level at all other on-pulse phases, that are also coincident with the phases when the predicted Beam2 dominates the $\gamma$-ray emission. This comparison suggests that, at any pulse phase, the polarisation angle is qualitatively correlated with the hemisphere dominating the emission (assuming similar emission geometries between X-ray and $\gamma$-ray).

   \begin{figure}
   \centering
   \includegraphics[width=.495\columnwidth]{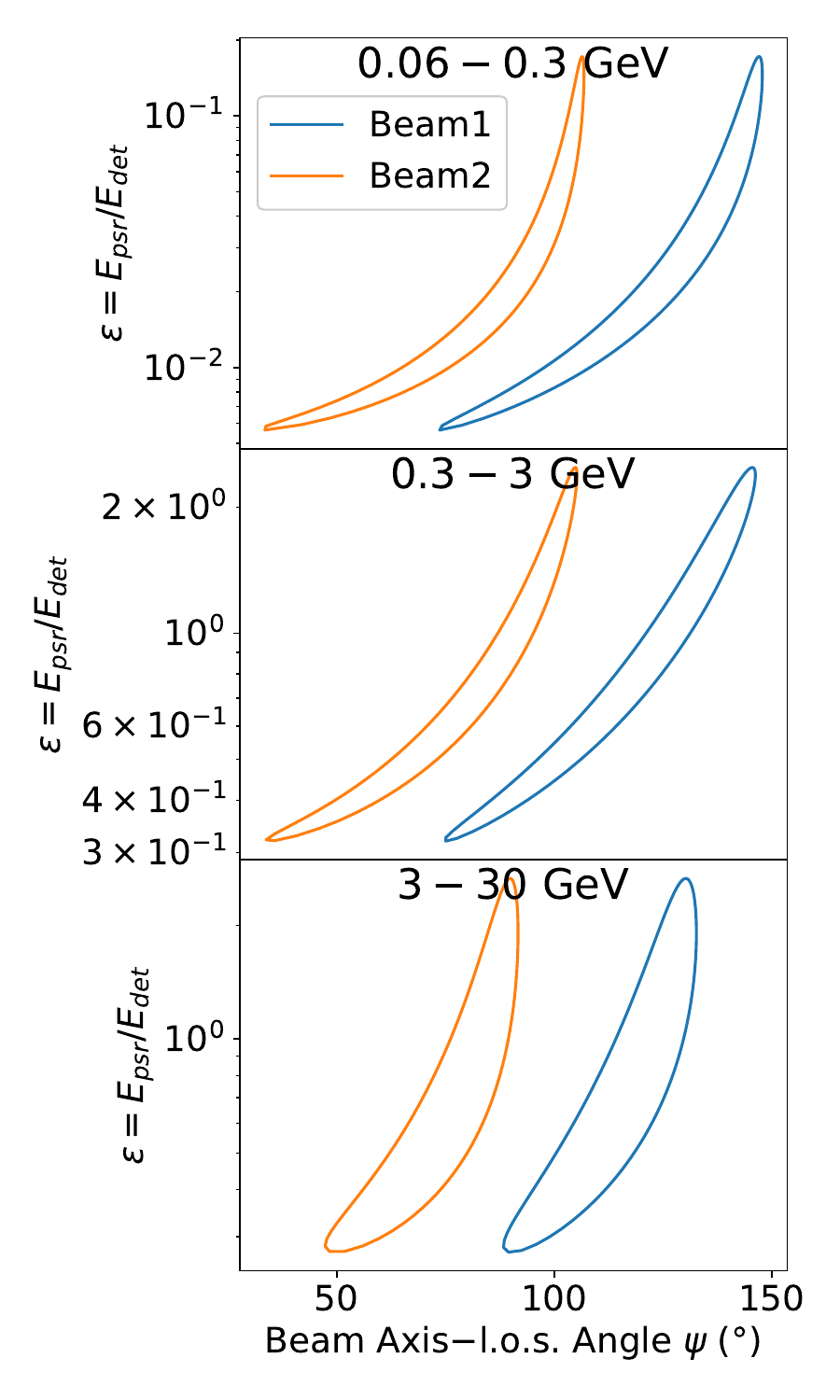}
   \includegraphics[width=.495\columnwidth]{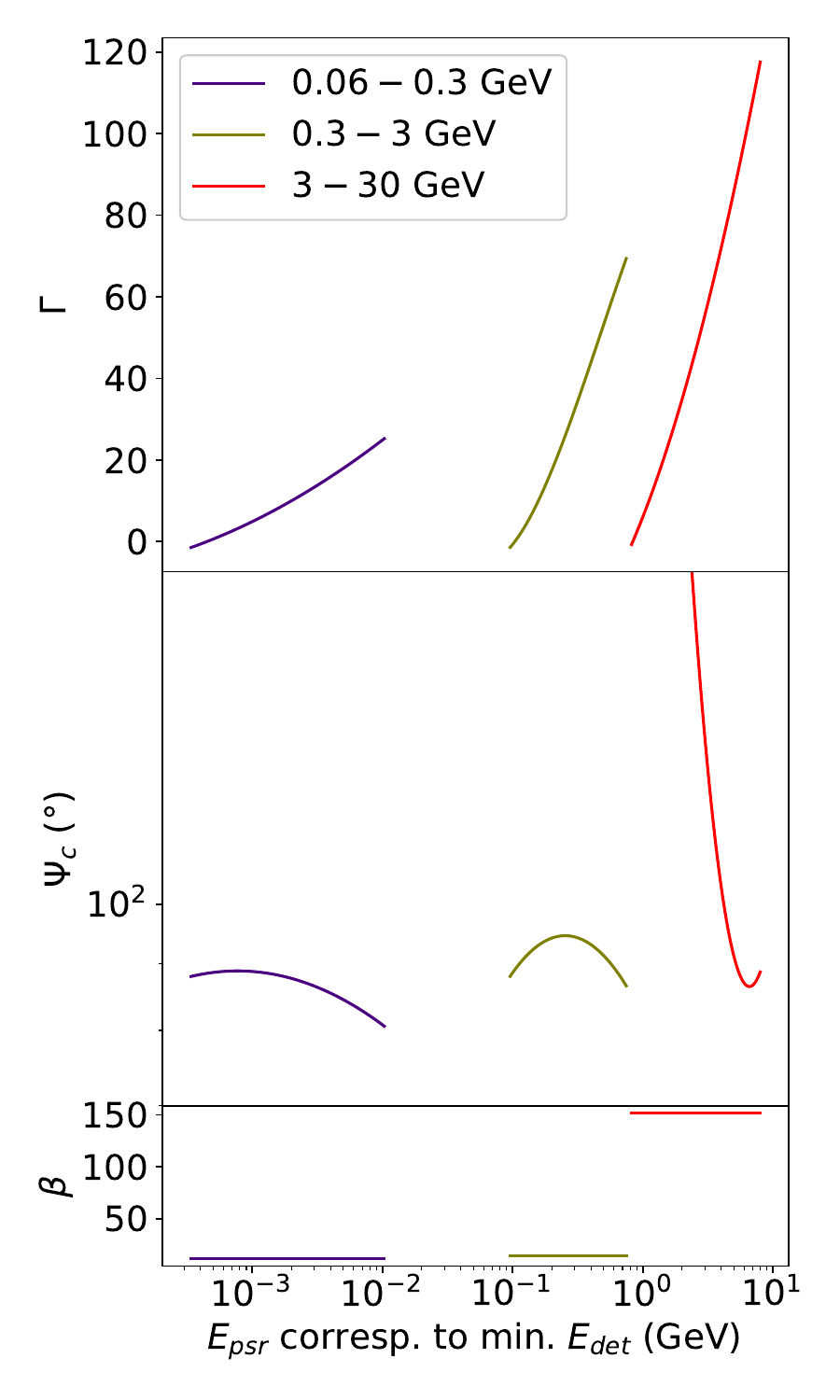}
   \caption{Systematic evaluations for results of the Crab pulsar. Left panels: Model-predicted contour of the observable outskirt of each beam, in an $(\psi,\varepsilon)$ space. Right panels: Model-predicted beam shape parameters and their energy-dependencies, referring to the minimum $E_{det}$ of each phaseogram. 
}
              \label{Crab_check}%
    \end{figure}

In addition, optical polarisation is qualitatively similar to X-ray polarisation \citep[Fig.~6 of][]{Wong_IXPE_2024}. Especially, the rise of the polarisation angle at P1’s TW is even more conspicuous in optical than in X-ray, further supporting our model prediction:  $\gamma$-rays at P1’s TW are predominantly emitted from a specific hemisphere, while those at  other on-pulse phases are predominantly emitted from the opposite hemisphere.

\subsection{Geminga pulsar}

   \begin{figure}
   \centering
   \includegraphics[width=.495\columnwidth]{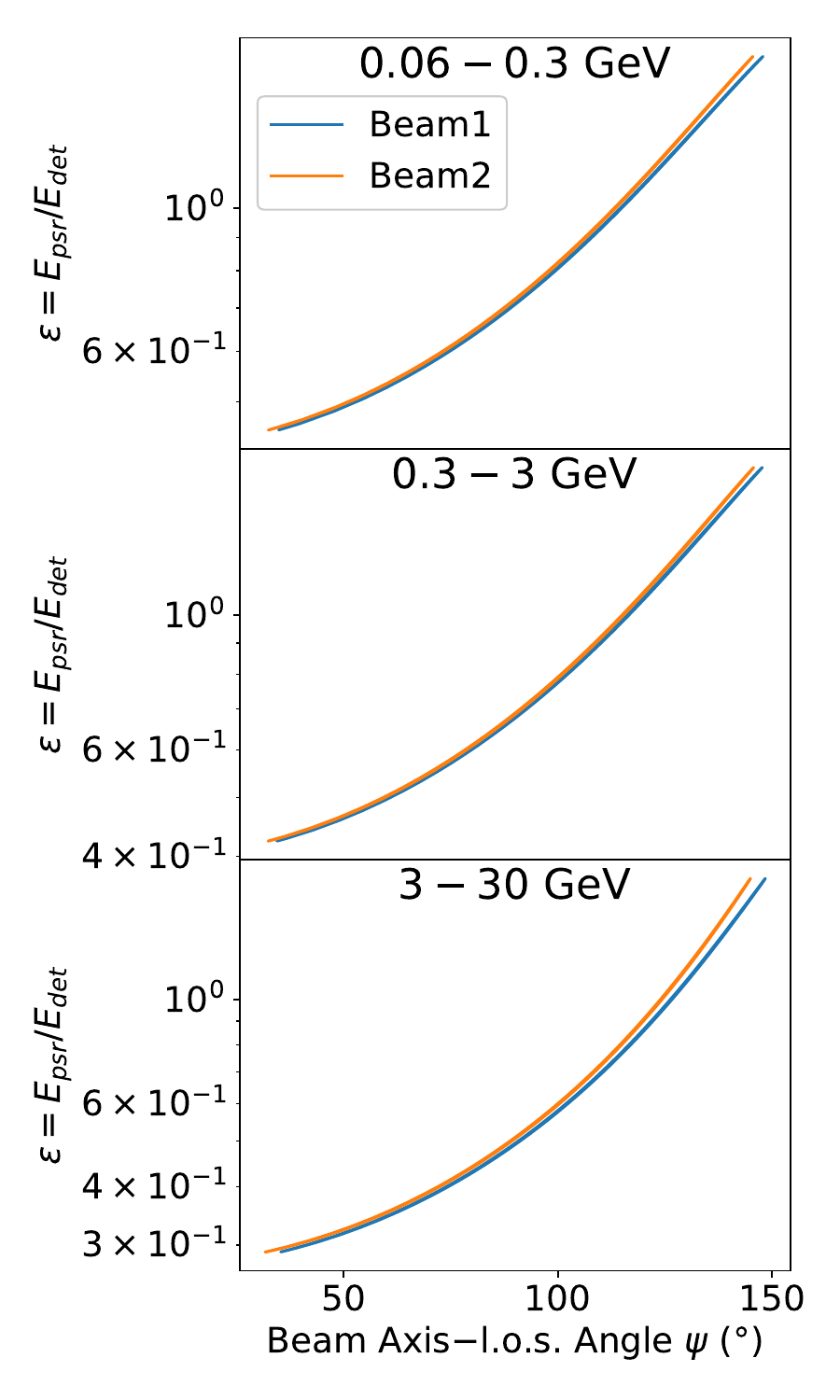}
   \includegraphics[width=.495\columnwidth]{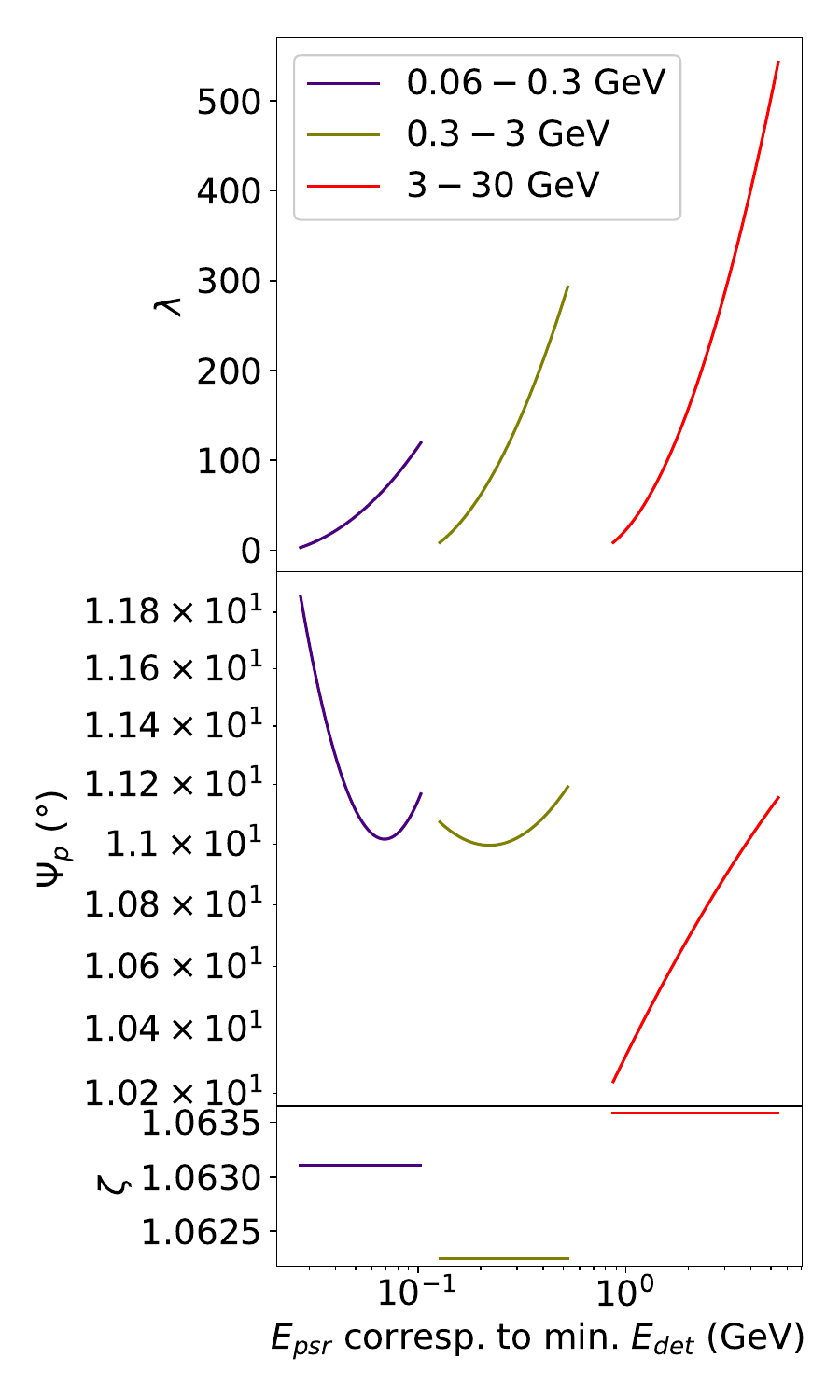}
   \caption{Systematic evaluations for results of the Geminga pulsar, where the more preferable model of MLP $\psi$-dependence is fitted. The structure of this figure is the same as in Fig.~\ref{Crab_check}.
}
              \label{Geminga_check}%
    \end{figure}

For the Geminga pulsar, our fittings to the 0.06$-$0.3~GeV and 0.3$-$3~GeV phaseograms predict approximately continuous trends of $\Psi_c$ (as shown in the middle-right panel of Fig.~\ref{Geminga_check}). However, there is a relatively drastic discontinuity between the trends of $\Psi_c$ predicted from the 0.3$-$3~GeV and 3$-$30~GeV phaseograms respectively. In a qualitative sense, the relatively drastic drop in $\Psi_c$ might also be relevant to a spectral shape transition and a mechanism transition at around 10~GeV, which are put forward by \citet{MAGIC_GemingaPSR_2020}. 

A number of features make the Geminga pulsar distinct from the Crab and Vela pulsars. Firstly, its strong preference on MLP $\psi$-dependence hints at a hollow conic structure of each beam. Secondly, both $\theta_B$ and $\theta_C$ are approximate to straight angles, indicating that the emission sites are almost on the equatorial plane and that the beam axes are almost horizontal. 

\PKHY{In an X-ray imaging analysis of the Geminga’s PWN that took the temporal variabilities into account, the inclination angle of its polar axis from our l.o.s. was constrained to be $<80\degree$ \citep{Hui_GemingaPWN_2017}. Our model-inferred $\Theta_A=56.7\degree$ is well within this upper limit.}

\subsubsection{Full-phase pulsed emission}

Interestingly, the un-pulsed count rate of the 0.06$-$0.3~GeV phaseogram is predicted to be much lower than the global minimum of the detected count rate, and the un-pulsed count rate predicted for the 0.3$-$3~GeV phaseogram is $\sim$69\% of the lowest detected count rate. These suggest that the Geminga pulsar does not have a genuine off-pulse phase, at least for tens to hundreds of MeV energies. Consequently, this finding poses a challenge to the traditional pulsar gating technique, which assumes the global minimum count rate of a phaseogram to be the so-called off-phase. For a full-phase pulsed emission, the traditional pulsar gating technique overestimates the un-pulsed (background) emission and underestimates the pulsed (source) emission.

Noticeably, the full-phase pulsed emission of the Geminga pulsar up to $\sim$2~GeV was also proposed by \citet{Abdo_GemingaPSR_2010}. Their phase-resolved spatial analyses and phase-resolved spectral analyses reached the same conclusion as ours: The so-called "off-pulse" phase range [0.70, 0.95] defined in \citet{MAGIC_GemingaPSR_2020} is not genuinely off. 

\subsection{Dragonfly pulsar}

   \begin{figure}
   \centering
   \includegraphics[width=.495\columnwidth]{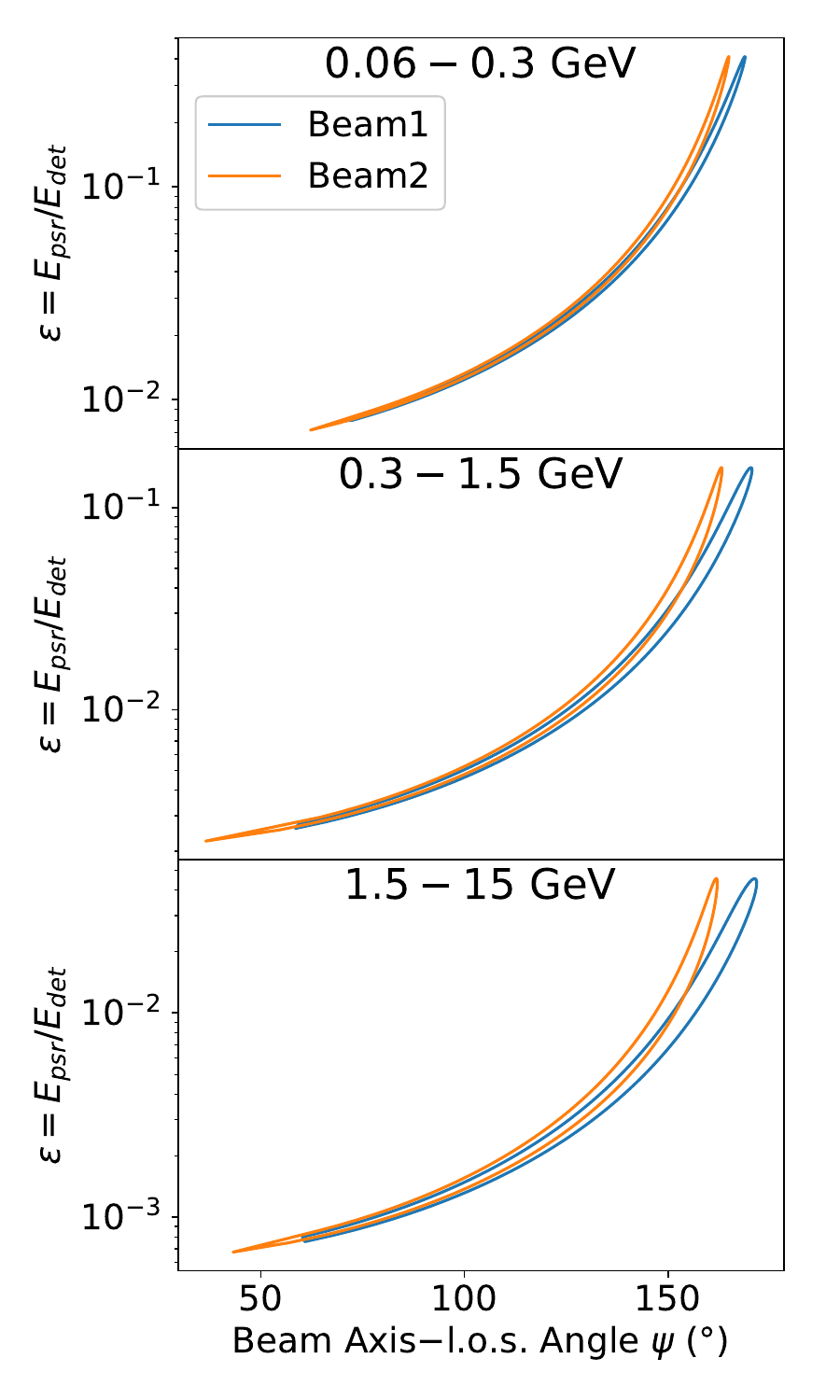}
   \includegraphics[width=.495\columnwidth]{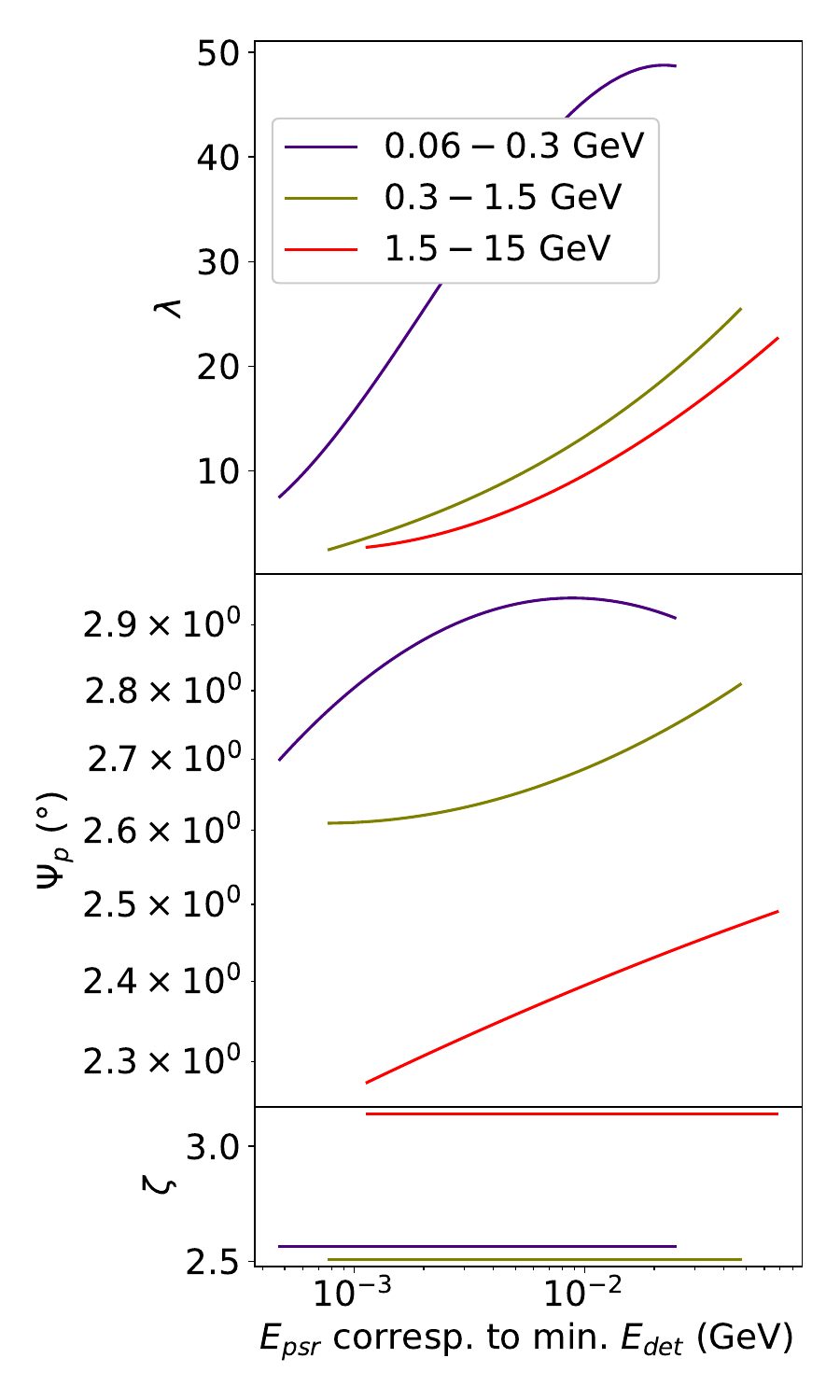}
   \caption{Systematic evaluations for results of the Dragonfly pulsar, where the more preferable model of MLP $\psi$-dependence is fitted. The structure of this figure is the same as in Fig.~\ref{Crab_check}.
}
              \label{Dragonfly_check}%
    \end{figure}

\PKHY{There are noteworthy similarities between the Dragonfly and Geminga pulsars. To be specific, a hollow conic structure of each beam is also favoured for the Dragonfly pulsar, and its nearly straight angles $\theta_C$ also indicate a pair of almost horizontal beam axes. The $\theta_B$ values obtained for the Dragonfly pulsar ($74.5\degree$--$82.6\degree$) are slightly smaller than those for the Geminga pulsar yet much larger than those for Crab and Vela pulsars, indicating that the emission sites are fairly close to the equatorial plane.}

\PKHY{Among the four pulsars in our case studies, the Dragonfly pulsar is found to have the largest $\Theta_A$, that is $77.0\degree$. Consistently, the radio and X-ray imaging analyses of the Dragonfly’s PWN also revealed that its polar axis has the largest inclination angle from our l.o.s. \citep[$85\degree$--$88\degree$;][]{VanEtten_DragonflyPWN_2008, Jin_DragonflyPWN_2023}. Their imaging results and our model-inferred value roughly agree within $11\degree$.}

\subsection{Vela pulsar}

   \begin{figure}
   \centering
   \includegraphics[width=.495\columnwidth]{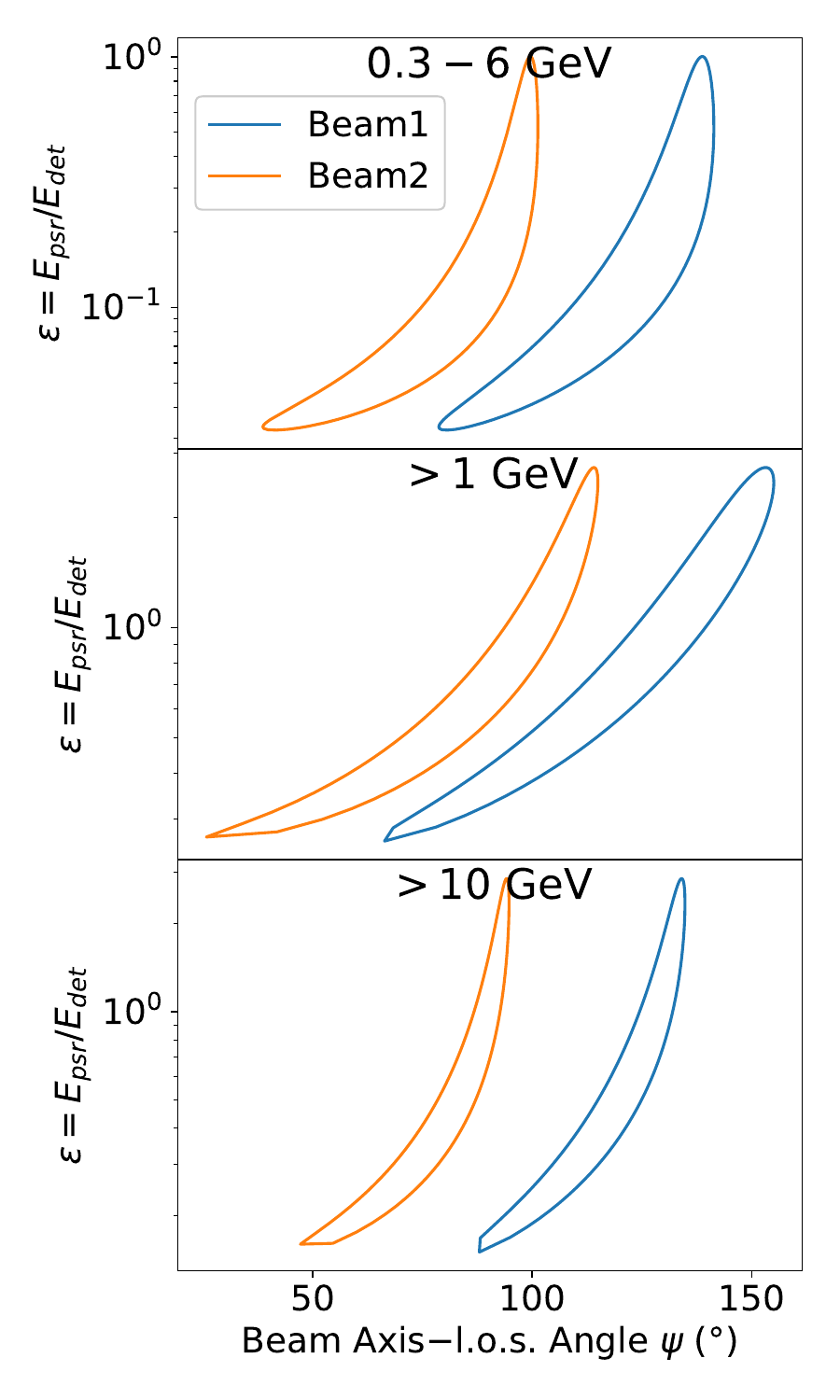}
   \includegraphics[width=.495\columnwidth]{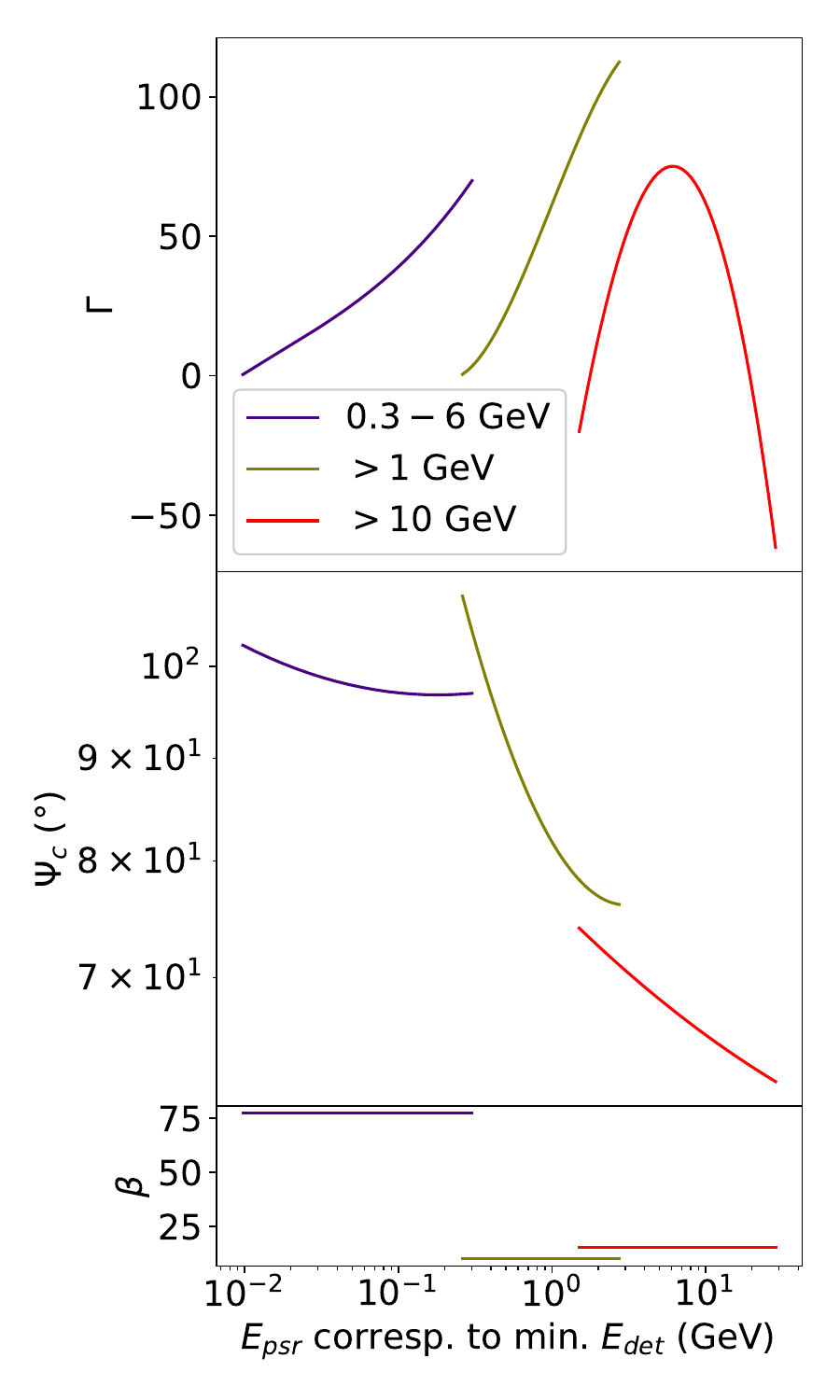}
   \caption{Systematic evaluations for results of the Vela pulsar. The structure of this figure is the same as in Fig.~\ref{Crab_check}.
}
              \label{Vela_check}%
    \end{figure}

The trends of $\Psi_c$ predicted from our fittings for the 0.3$-$6~GeV, $>1~$GeV and $>10~$GeV phaseograms of the Vela pulsar are roughly connected with one another (as shown in the middle-right panel of Fig.~\ref{Vela_check}). The wind-current-sheet model established by \citet{HESS_Vela_2023} argues that the transition from a synchrotron/curvature-dominated regime to an inverse-Compton-dominated regime occurs at a much higher energy of $>0.1~$TeV (this transition manifests itself as a segregation of two spectral components). Therefore, this transition does not seem to be relevant to the data sample adopted in this work. In turn, our findings serve as further, indirect evidence that a single radiation mechanism (synchrotron/curvature radiation) could reproduce \emph{Fermi}-LAT $\gamma$-ray phaseograms of the Vela pulsar from hundred MeV to tens of GeV.

The polar axis of the Vela PWN's equatorial torus was constrained to be inclined at $63\degree$--$64\degree$ from our l.o.s. \citep[through the imaging analysis;][]{Ng_Tori_2004}. It is comforting to note that this measurement roughly agrees with our determined $\Theta_A=70.0\degree$ for the Vela pulsar.

\section{Evaluation and potential improvements}

Despite the qualitative match between data and simulation for \PKHY{twelve phaseograms of four pulsars}, we evaluate the predictive power of our prototypical toy model, based on Fig.~\ref{Crab_check}--\ref{Vela_check}. According to the left panels, the photons with a specific $\varepsilon$ are always detected at four specific $\psi$. Exceptionally, the Geminga pulsar’s model predicts a much narrower range of $\psi$ for any $\varepsilon$. From a statistical point of view, our attempted PLSEC and MLP functions, each of which consists of three energy-dependent variables plus one locally uniform$^{\ref{LocUni}}$ variable, are sufficient to approximate the $\psi$-dependence of $C$ at a fixed $\varepsilon$.

For each pulsar, the fittings of different phaseograms predict discontinuous and discrepant trends of beam shape parameters (as shown in right panels of Fig.~\ref{Crab_check}--\ref{Vela_check}). On the other hand, the different locations, vertical extensions and speeds of emission sites ($d$, $\theta_B$, $t_0$, $\sigma_{ver}$ and $\gamma_{Lor}$) determined from different phaseograms of the same pulsar indicate that photons detected at different $E_{det}$ are actually emitted from different (but overlapping) populations of particles. Realistically, even for a specific phase, the mapping between $E_{psr}$ and $E_{det}$ has a many-to-many relation. In other words, photons with the same $E_{psr}$ are Doppler-shifted to a range of $E_{det}$ values, and a single $E_{det}$ corresponds to a range of $E_{psr}$ values. From this perspective,  the trends of beam shape parameters inferred from different phaseograms of the same pulsar do not need to be consistent or connected. Furthermore, the spectral shape transition  at higher energies could also come into the play for the discontinuity and discrepancy in the trends of beam shape parameters.

Nevertheless, those very large values obtained for $\Gamma$ of PLSEC and $\lambda$ of MLP, as well as  their energy dependencies,  deserve our deeper review.  It is unclear what kinds of astrophysical processes could cause such extremely large values. From a data-driven perspective, an extremely energy-dependent beam shape is a necessary cost for the Doppler effect to reproduce the Bridge/P3 emission of a pulsar, especially under our assumptions that a pulsar has a north-south symmetric configuration, and that each hemisphere only harbours a single  emission site emitting a circular beam with a locally uniform$^{\ref{LocUni}}$ axis. Our neglect of the \PKHY{location-dependent variations of sub-beam properties} could have further increased the cost. Additionally, we recall that our observations always cover a limited outskirt region of a beam, resulting in poorer constraints on the prefactor $N$, $\Gamma$ of PLSEC and $\lambda$ of MLP.

It is worthwhile to mention that the beam axis angles $\theta_C$ and $\theta_Q$ seem to be slightly energy-dependent as well. No matter whether the measured energy-dependent shifting of a beam axis is a physical phenomenon or a statistical/systematic fluctuation, our assumption of a locally uniform$^{\ref{LocUni}}$ axis direction for a phaseogram could be a deficiency, and its neglected energy-dependence could have been allocated to the beam shape and the prefactor $N$.

\REFE{There is limited compatibility between our existing model and the wind-current-sheet model (despite the resemblances pinpointed in \S\ref{WindModels}). Our existing model considers only one “dummy” emission site (not even azimuthally extended) and one “dummy” beam per hemisphere. These are  insufficient to capture the geometrical complexity of a wind current sheet \citep[e.g. Fig.~4 of][]{HESS_Vela_2023}.}

In a nutshell, we acknowledge  considerable room for boosting the predictive power of our toy model while maintaining the north-south symmetry. In the following, we outline some improvable aspects of our toy model, which are useful for us to establish an advanced toy model in a follow-up work.

\subsection{Asymmetry of each beam and its energy-dependence}\label{Asym_beam}

One possibility is that the actual beam shape is significantly asymmetric and our approximation of it as a circular beam leads to mis-measurements of its beam axis direction. What is more, its energy-dependent level of asymmetry could make an illusion of energy-dependent $\theta_C$ and $\theta_Q$. We speculate that taking the energy-dependent asymmetry of a beam into account could  partially reconcile the misalignment of beam axes among different energies.

\subsection{Energy-dependent direction of each beam axis}\label{E_dep_loc}

If the beam axis directions at different energies remain significantly discrepant, then we may parametrise the energy-dependent shifting of $\theta_C$ and $\theta_Q$ as well. A combination of an energy-dependent beam axis and energy-dependent asymmetry could, to a certain extent, serve as a substitute for the extreme energy-dependence of $\Gamma$, $\lambda$ and $N$.

\subsection{Multiple emission sites and multiple misaligned beams per hemisphere}\label{MultipleComponents}

It is also meaningful to test an advanced toy model that treats the Bridge/P3 emission as a \REFE{second} pair of components (i.e. Beam3 and Beam4) radiated from a \REFE{second} pair of emission regions. This is particularly important for the Vela pulsar because of its arch-like Bridge. In this way, we expect the Doppler-boosted tail and/or sub-spikes of each component to be less pronounced. \REFESEC{Such an advanced model will also enable us to re-assess the presence of the hypothetical “Bridge2/P4” unbiasedly, and to measure the fractional contributions of the northern and southern hemispheres to Vela pulsar's  emissions more accurately.}

If each hemisphere of a pulsar harbours multiple  emission sites emitting multiple  beams of misaligned axes, then a "dummy" emission site and a "dummy" beam considered in our prototypical toy model could be interpreted as ensembles of multiple components. Those inferred  properties are, in some sense,  the weighted averages of multiple components. An estimated  extension  in this work could be associated with the separations among multiple components in addition to their respective sizes.

\REFE{Multiple  emission sites  could have either the same or different essences/mechanisms. What is more, they could be either spatially connected or discontinuous. Noteworthily, the complicated wind-current-sheet region \citep[e.g. Fig.~4 of][]{HESS_Vela_2023}  could be more accurately approximated as multiple pairs of “dummy” sites emitting multiple pairs of “dummy” beams. Therefore, incorporating this idea (multiple components per hemisphere) into our toy model can further increase its compatibility with the wind-current-sheet model.}

In order to reduce the inter-component correlation, we may start from  a \PKHY{relatively simple} scenario of double unextended sites and double circular beams per hemisphere, with energy-independent site locations and beam axes. In this scenario, the energy-dependent proportion of the two component beams would shape the energy-dependent asymmetry and energy-dependent peak direction of the resultant combined beam at one hemisphere. Therefore, this idea can be considered inclusive of the ideas of \S\ref{Asym_beam} and \S\ref{E_dep_loc}. Indeed, a combination of two beams with misaligned axes could allow a much wider diversity of the resultant beam morphology, greatly reducing the cost for the Doppler effect to reproduce the Bridge/P3 emission of a pulsar.

\section{Summary}

This work establishes a prototype of a toy model, whose innovative feature is that it is independent of emission sites and radiation mechanisms. Our prototypical toy model  assumes a north-south symmetric configuration, and considers a single emission site and a single, circular beam per hemisphere. It takes into account the emission geometry, Doppler shifts and time delays, as well as the directional dependence and spectral dependence of a beam’s count rate. It, at least qualitatively, reproduces the observed pulse features of Crab, Geminga, Dragonfly and Vela pulsars. These include: The LW--TW asymmetry of P1 and P2; The valley-floor shape of Crab’s, Geminga’s and Dragonfly’s Bridges; The “bump” at Vela’s Bridge.

Our fitting results preliminarily suggest that  the third components of these four pulsars could  be interpreted as Doppler-boosted tails of Beam2. It is also preliminarily hinted that the combination of our l.o.s., the emission geometry and the energy-dependent beam shape disfavours the detection of a hypothetical “Bridge2/P4” \PKHY{(i.e. a hypothetical counterpart of Bridge/P3 emitted from the opposite hemisphere)}. Nonetheless, these two claims need to be further validated or interrogated in a follow-up work. Besides, our findings of \PKHY{$0.88<d\sin{\theta_B}/R_{LC}<1.31$} and $90\degree<\theta_Q<115\degree$, once confirmed in a follow-up work, will give further endorsement to wind models \citep[e.g.,][]{Aharonian_wind_2012,HESS_Vela_2023}. Moreover, for the Crab pulsar, a preliminary qualitative correlation between the X-ray polarisation angle and the dominating $\gamma$-ray pulse component is awaiting our further crosscheck with an advanced toy model.

The predictive power of our model is currently somewhat limited, owing to an unusually high requirement for the energy dependence of a beam shape. Depending on the data amount and quality, we will introduce further geometrical complexities to our toy model. We speculate that these geometrical complexities will, to some extent, replace the extreme energy-dependencies of the beam shapes inferred with the current model. After our toy model evolves into a more mature form, we will interpret the essence and radiation mechanism of each emission site more robustly, based on our inferred properties. Ultimately, we will also concretise the connection between our toy model and commonly proposed astrophysical theories.

\REFE{Extending our case studies to other $\gamma$-ray pulsars is also an indispensable part of our follow-up work. For fainter ones and for those showing only one peak on a phaseogram, the correlations and degeneracies among parameters become more serious even when we fit with our prototypical toy model. These could be compensated by reducing some free variables (e.g. removing higher-order coefficients of Eqn.~\ref{N_delta}--\ref{Psi_c_delta}~\&~\ref{lambda_delta}--\ref{Psi_p_delta}, assigning $\beta$ or $\zeta$ a globally uniform distribution).}

\section*{Acknowledgements}

DK acknowledges support by JSPS KAKENHI Grant Number 23H04891.
We thank Diego F. Torres, Marcos López Moya, Tsunefumi Mizuno and Matthew Kerr for very useful discussion. We also thank the anonymous referee for the fruitful comments.

\section*{Data availability}

The \emph{Fermi}-LAT photon data involved in this paper can be accessed from the Fermi-LAT data server: \url{https://fermi.gsfc.nasa.gov/cgi-bin/ssc/LAT/LATDataQuery.cgi}.




\begin{thebibliography}{}
\makeatletter
\relax
\def\mn@urlcharsother{\let\do\@makeother \do\$\do\&\do\#\do\^\do\_\do\%\do\~}
\def\mn@doi{\begingroup\mn@urlcharsother \@ifnextchar [ {\mn@doi@} {\mn@doi@[]}}
\def\mn@doi@[#1]#2{\def\@tempa{#1}\ifx\@tempa\@empty \href {http://dx.doi.org/#2} {doi:#2}\else \href {http://dx.doi.org/#2} {#1}\fi \endgroup}
\def\mn@eprint#1#2{\mn@eprint@#1:#2::\@nil}
\def\mn@eprint@arXiv#1{\href {http://arxiv.org/abs/#1} {{\tt arXiv:#1}}}
\def\mn@eprint@dblp#1{\href {http://dblp.uni-trier.de/rec/bibtex/#1.xml} {dblp:#1}}
\def\mn@eprint@#1:#2:#3:#4\@nil{\def\@tempa {#1}\def\@tempb {#2}\def\@tempc {#3}\ifx \@tempc \@empty \let \@tempc \@tempb \let \@tempb \@tempa \fi \ifx \@tempb \@empty \def\@tempb {arXiv}\fi \@ifundefined {mn@eprint@\@tempb}{\@tempb:\@tempc}{\expandafter \expandafter \csname mn@eprint@\@tempb\endcsname \expandafter{\@tempc}}}

\bibitem[\protect\citeauthoryear{{Abdo} et~al.,}{{Abdo} et~al.}{2010a}]{Abdo_Crab_2010}
{Abdo} A.~A.,  et~al., 2010a, \mn@doi [\apj] {10.1088/0004-637X/708/2/1254}, \href {https://ui.adsabs.harvard.edu/abs/2010ApJ...708.1254A} {708, 1254}

\bibitem[\protect\citeauthoryear{{Abdo} et~al.,}{{Abdo} et~al.}{2010b}]{Abdo2010_VelaPSR}
{Abdo} A.~A.,  et~al., 2010b, \mn@doi [\apj] {10.1088/0004-637X/713/1/154}, \href {https://ui.adsabs.harvard.edu/abs/2010ApJ...713..154A} {713, 154}

\bibitem[\protect\citeauthoryear{{Abdo} et~al.,}{{Abdo} et~al.}{2010c}]{Abdo_GemingaPSR_2010}
{Abdo} A.~A.,  et~al., 2010c, \mn@doi [\apj] {10.1088/0004-637X/720/1/272}, \href {https://ui.adsabs.harvard.edu/abs/2010ApJ...720..272A} {720, 272}

\bibitem[\protect\citeauthoryear{{Abe} et~al.,}{{Abe} et~al.}{2024}]{CTAO-LST_CrabPSR_2024}
{Abe} K.,  et~al., 2024, \mn@doi [\aap] {10.1051/0004-6361/202450059}, \href {https://ui.adsabs.harvard.edu/abs/2024A&A...690A.167A} {690, A167}

\bibitem[\protect\citeauthoryear{{Abe} et~al.,}{{Abe} et~al.}{2025}]{CTAO-LST_GemingaPSR_2025}
{Abe} K.,  et~al., 2025, \mn@doi [\aap] {10.1051/0004-6361/202554350}, \href {https://ui.adsabs.harvard.edu/abs/2025A&A...698A.283A} {698, A283}

\bibitem[\protect\citeauthoryear{{Aharonian}, {Bogovalov}  \& {Khangulyan}}{{Aharonian} et~al.}{2012}]{Aharonian_wind_2012}
{Aharonian} F.~A.,  {Bogovalov} S.~V.,   {Khangulyan} D.,  2012, \mn@doi [\nat] {10.1038/nature10793}, \href {https://ui.adsabs.harvard.edu/abs/2012Natur.482..507A} {482, 507}

\bibitem[\protect\citeauthoryear{{Ahnen} et~al.,}{{Ahnen} et~al.}{2016}]{Ahnen2016}
{Ahnen} M.~L.,  et~al., 2016, \mn@doi [\aap] {10.1051/0004-6361/201527722}, \href {https://ui.adsabs.harvard.edu/abs/2016A&A...591A.138A} {591, A138}

\bibitem[\protect\citeauthoryear{{Aleksi{\'c}} et~al.,}{{Aleksi{\'c}} et~al.}{2012}]{Aleksic_Gap_2012}
{Aleksi{\'c}} J.,  et~al., 2012, \mn@doi [\aap] {10.1051/0004-6361/201118166}, \href {https://ui.adsabs.harvard.edu/abs/2012A&A...540A..69A} {540, A69}

\bibitem[\protect\citeauthoryear{{Aleksi{\'c}} et~al.,}{{Aleksi{\'c}} et~al.}{2014}]{Aleksic_BD_2014}
{Aleksi{\'c}} J.,  et~al., 2014, \mn@doi [\aap] {10.1051/0004-6361/201423664}, \href {https://ui.adsabs.harvard.edu/abs/2014A&A...565L..12A} {565, L12}

\bibitem[\protect\citeauthoryear{{Ansoldi} et~al.,}{{Ansoldi} et~al.}{2016}]{ansoldi_teraelectronvolt_2016}
{Ansoldi} S.,  et~al., 2016, \mn@doi [\aap] {10.1051/0004-6361/201526853}, \href {https://ui.adsabs.harvard.edu/abs/2016A&A...585A.133A} {585, A133}

\bibitem[\protect\citeauthoryear{{Arons} \& {Scharlemann}}{{Arons} \& {Scharlemann}}{1979}]{1979ApJ...231..854A}
{Arons} J.,  {Scharlemann} E.~T.,  1979, \mn@doi [\apj] {10.1086/157250}, \href {https://ui.adsabs.harvard.edu/abs/1979ApJ...231..854A} {231, 854}

\bibitem[\protect\citeauthoryear{{Bai} \& {Spitkovsky}}{{Bai} \& {Spitkovsky}}{2010}]{Bai_caustic_2010}
{Bai} X.-N.,  {Spitkovsky} A.,  2010, \mn@doi [\apj] {10.1088/0004-637X/715/2/1282}, \href {https://ui.adsabs.harvard.edu/abs/2010ApJ...715.1282B} {715, 1282}

\bibitem[\protect\citeauthoryear{{Cheng}, {Ho}  \& {Ruderman}}{{Cheng} et~al.}{1986}]{1986ApJ...300..500C}
{Cheng} K.~S.,  {Ho} C.,   {Ruderman} M.,  1986, \mn@doi [\apj] {10.1086/163829}, \href {https://ui.adsabs.harvard.edu/abs/1986ApJ...300..500C} {300, 500}

\bibitem[\protect\citeauthoryear{{Deutsch}}{{Deutsch}}{1955}]{1955AnAp...18....1D}
{Deutsch} A.~J.,  1955, Annales d'Astrophysique, \href {https://ui.adsabs.harvard.edu/abs/1955AnAp...18....1D} {18, 1}

\bibitem[\protect\citeauthoryear{{Dyks} \& {Rudak}}{{Dyks} \& {Rudak}}{2003}]{Dyks_caustic_2003}
{Dyks} J.,  {Rudak} B.,  2003, \mn@doi [\apj] {10.1086/379052}, \href {https://ui.adsabs.harvard.edu/abs/2003ApJ...598.1201D} {598, 1201}

\bibitem[\protect\citeauthoryear{{Fierro}, {Michelson}, {Nolan}  \& {Thompson}}{{Fierro} et~al.}{1998}]{Fierro_EGRET_1998}
{Fierro} J.~M.,  {Michelson} P.~F.,  {Nolan} P.~L.,   {Thompson} D.~J.,  1998, \mn@doi [\apj] {10.1086/305219}, \href {https://ui.adsabs.harvard.edu/abs/1998ApJ...494..734F} {494, 734}

\bibitem[\protect\citeauthoryear{{Gold}}{{Gold}}{1968}]{1968Natur.218..731G}
{Gold} T.,  1968, \nat

\bibitem[\protect\citeauthoryear{{Goldreich} \& {Julian}}{{Goldreich} \& {Julian}}{1969}]{1969ApJ...157..869G}
{Goldreich} P.,  {Julian} W.~H.,  1969, \mn@doi [\apj] {10.1086/150119}, \href {https://ui.adsabs.harvard.edu/abs/1969ApJ...157..869G} {157, 869}

\bibitem[\protect\citeauthoryear{{Gonz{\'a}lez-Caniulef}, {Heyl}, {Fabiani}, {Soffitta}, {Costa}, {Bucciantini}, {Kirmizibayrak}  \& {Xie}}{{Gonz{\'a}lez-Caniulef} et~al.}{2025}]{Gonzalez2025}
{Gonz{\'a}lez-Caniulef} D.,  {Heyl} J.,  {Fabiani} S.,  {Soffitta} P.,  {Costa} E.,  {Bucciantini} N.,  {Kirmizibayrak} D.,   {Xie} F.,  2025, \mn@doi [\aap] {10.1051/0004-6361/202451815}, \href {https://ui.adsabs.harvard.edu/abs/2025A&A...693A.152G} {693, A152}

\bibitem[\protect\citeauthoryear{{H.~E.~S.~S. Collaboration} et~al.,}{{H.~E.~S.~S. Collaboration} et~al.}{2018}]{HESS_Vela_2018}
{H.~E.~S.~S. Collaboration} et~al., 2018, \mn@doi [\aap] {10.1051/0004-6361/201732153}, \href {https://ui.adsabs.harvard.edu/abs/2018A&A...620A..66H} {620, A66}

\bibitem[\protect\citeauthoryear{{H.~E.~S.~S. Collaboration} et~al.,}{{H.~E.~S.~S. Collaboration} et~al.}{2023}]{HESS_Vela_2023}
{H.~E.~S.~S. Collaboration} et~al., 2023, \mn@doi [Nature Astronomy] {10.1038/s41550-023-02052-3}, \href {https://ui.adsabs.harvard.edu/abs/2023NatAs...7.1341H} {7, 1341}

\bibitem[\protect\citeauthoryear{{Harding}, {Venter}  \& {Kalapotharakos}}{{Harding} et~al.}{2021}]{Harding_model_2021}
{Harding} A.~K.,  {Venter} C.,   {Kalapotharakos} C.,  2021, \mn@doi [\apj] {10.3847/1538-4357/ac3084}, \href {https://ui.adsabs.harvard.edu/abs/2021ApJ...923..194H} {923, 194}

\bibitem[\protect\citeauthoryear{{Hewish}, {Bell}, {Pilkington}, {Scott}  \& {Collins}}{{Hewish} et~al.}{1968}]{1968Natur.217..709H}
{Hewish} A.,  {Bell} S.~J.,  {Pilkington} J.~D.~H.,  {Scott} P.~F.,   {Collins} R.~A.,  1968, \mn@doi [\nat] {10.1038/217709a0}, \href {https://ui.adsabs.harvard.edu/abs/1968Natur.217..709H} {217, 709}

\bibitem[\protect\citeauthoryear{{Hui}, {Lee}, {Kong}, {Tam}, {Takata}, {Cheng}  \& {Ryu}}{{Hui} et~al.}{2017}]{Hui_GemingaPWN_2017}
{Hui} C.~Y.,  {Lee} J.,  {Kong} A.~K.~H.,  {Tam} P.~H.~T.,  {Takata} J.,  {Cheng} K.~S.,   {Ryu} D.,  2017, \mn@doi [\apj] {10.3847/1538-4357/aa862c}, \href {https://ui.adsabs.harvard.edu/abs/2017ApJ...846..116H} {846, 116}

\bibitem[\protect\citeauthoryear{{Jin}, {Ng}, {Roberts}  \& {Li}}{{Jin} et~al.}{2023}]{Jin_DragonflyPWN_2023}
{Jin} R.,  {Ng} C.~Y.,  {Roberts} M. S.~E.,   {Li} K.-L.,  2023, \mn@doi [\apj] {10.3847/1538-4357/aca656}, \href {https://ui.adsabs.harvard.edu/abs/2023ApJ...942..100J} {942, 100}

\bibitem[\protect\citeauthoryear{{Johnson} \& {Teller}}{{Johnson} \& {Teller}}{1982}]{Johnson1982}
{Johnson} M.~H.,  {Teller} E.,  1982, \mn@doi [Proceedings of the National Academy of Science] {10.1073/pnas.79.4.1340}, \href {https://ui.adsabs.harvard.edu/abs/1982PNAS...79.1340J} {79, 1340}

\bibitem[\protect\citeauthoryear{{Kargaltsev}, {Hare}  \& {Lange}}{{Kargaltsev} et~al.}{2023}]{Kargaltsev2023}
{Kargaltsev} O.,  {Hare} J.,   {Lange} A.,  2023, \mn@doi [arXiv e-prints] {10.48550/arXiv.2312.03198}, \href {https://ui.adsabs.harvard.edu/abs/2023arXiv231203198K} {p. arXiv:2312.03198}

\bibitem[\protect\citeauthoryear{{Leung}, {Takata}, {Ng}, {Kong}, {Tam}, {Hui}  \& {Cheng}}{{Leung} et~al.}{2014}]{Leung2014}
{Leung} G. C.~K.,  {Takata} J.,  {Ng} C.~W.,  {Kong} A.~K.~H.,  {Tam} P.~H.~T.,  {Hui} C.~Y.,   {Cheng} K.~S.,  2014, \mn@doi [\apjl] {10.1088/2041-8205/797/2/L13}, \href {https://ui.adsabs.harvard.edu/abs/2014ApJ...797L..13L} {797, L13}

\bibitem[\protect\citeauthoryear{{Luo} et~al.,}{{Luo} et~al.}{2019}]{Luo2019}
{Luo} J.,  et~al., 2019, {PINT: High-precision pulsar timing analysis package}, Astrophysics Source Code Library, record ascl:1902.007

\bibitem[\protect\citeauthoryear{{Luo} et~al.,}{{Luo} et~al.}{2021}]{Luo2021}
{Luo} J.,  et~al., 2021, \mn@doi [\apj] {10.3847/1538-4357/abe62f}, \href {https://ui.adsabs.harvard.edu/abs/2021ApJ...911...45L} {911, 45}

\bibitem[\protect\citeauthoryear{{MAGIC Collaboration} et~al.,}{{MAGIC Collaboration} et~al.}{2020}]{MAGIC_GemingaPSR_2020}
{MAGIC Collaboration} et~al., 2020, \mn@doi [\aap] {10.1051/0004-6361/202039131}, \href {https://ui.adsabs.harvard.edu/abs/2020A&A...643L..14M} {643, L14}

\bibitem[\protect\citeauthoryear{{Ng} \& {Romani}}{{Ng} \& {Romani}}{2004}]{Ng_Tori_2004}
{Ng} C.~Y.,  {Romani} R.~W.,  2004, \mn@doi [\apj] {10.1086/380486}, \href {https://ui.adsabs.harvard.edu/abs/2004ApJ...601..479N} {601, 479}

\bibitem[\protect\citeauthoryear{{P{\'e}tri}}{{P{\'e}tri}}{2015}]{Petri_caustic_2015}
{P{\'e}tri} J.,  2015, \mn@doi [\aap] {10.1051/0004-6361/201424289}, \href {https://ui.adsabs.harvard.edu/abs/2015A&A...574A..51P} {574, A51}

\bibitem[\protect\citeauthoryear{{Ruderman} \& {Sutherland}}{{Ruderman} \& {Sutherland}}{1975}]{1975ApJ...196...51R}
{Ruderman} M.~A.,  {Sutherland} P.~G.,  1975, \mn@doi [\apj] {10.1086/153393}, \href {https://ui.adsabs.harvard.edu/abs/1975ApJ...196...51R} {196, 51}

\bibitem[\protect\citeauthoryear{{S{\l}owikowska}, {Kanbach}, {Kramer}  \& {Stefanescu}}{{S{\l}owikowska} et~al.}{2009}]{Slowikowska2009}
{S{\l}owikowska} A.,  {Kanbach} G.,  {Kramer} M.,   {Stefanescu} A.,  2009, \mn@doi [\mnras] {10.1111/j.1365-2966.2009.14935.x}, \href {https://ui.adsabs.harvard.edu/abs/2009MNRAS.397..103S} {397, 103}

\bibitem[\protect\citeauthoryear{{Smith} et~al.,}{{Smith} et~al.}{2023}]{Smith_3PC_2023}
{Smith} D.~A.,  et~al., 2023, \mn@doi [\apj] {10.3847/1538-4357/acee67}, \href {https://ui.adsabs.harvard.edu/abs/2023ApJ...958..191S} {958, 191}

\bibitem[\protect\citeauthoryear{{Sturrock}}{{Sturrock}}{1971}]{1971ApJ...164..529S}
{Sturrock} P.~A.,  1971, \mn@doi [\apj] {10.1086/150865}, \href {https://ui.adsabs.harvard.edu/abs/1971ApJ...164..529S} {164, 529}

\bibitem[\protect\citeauthoryear{{Van Etten}, {Romani}  \& {Ng}}{{Van Etten} et~al.}{2008}]{VanEtten_DragonflyPWN_2008}
{Van Etten} A.,  {Romani} R.~W.,   {Ng} C.~Y.,  2008, \mn@doi [\apj] {10.1086/587865}, \href {https://ui.adsabs.harvard.edu/abs/2008ApJ...680.1417V} {680, 1417}

\bibitem[\protect\citeauthoryear{Wang}{Wang}{2025}]{Wang_thesis_2025}
Wang Y.,  2025, Master thesis, Ludwig-Maximilians-Universit\"at M\"unchen

\bibitem[\protect\citeauthoryear{{Wong} et~al.,}{{Wong} et~al.}{2024}]{Wong_IXPE_2024}
{Wong} J.,  et~al., 2024, \mn@doi [\apj] {10.3847/1538-4357/ad6309}, \href {https://ui.adsabs.harvard.edu/abs/2024ApJ...973..172W} {973, 172}

\bibitem[\protect\citeauthoryear{{Yeung}}{{Yeung}}{2020}]{Yeung_CrabPSR_2020}
{Yeung} P. K.~H.,  2020, \mn@doi [\aap] {10.1051/0004-6361/202038166}, \href {https://ui.adsabs.harvard.edu/abs/2020A&A...640A..43Y} {640, A43}

\makeatother
\end{thebibliography}
\input{output.bbl}




\appendix

\section{Crosschecking results}\label{ApxGeminga}

\begin{figure*}
   \centering
   \includegraphics[width=.33\textwidth]{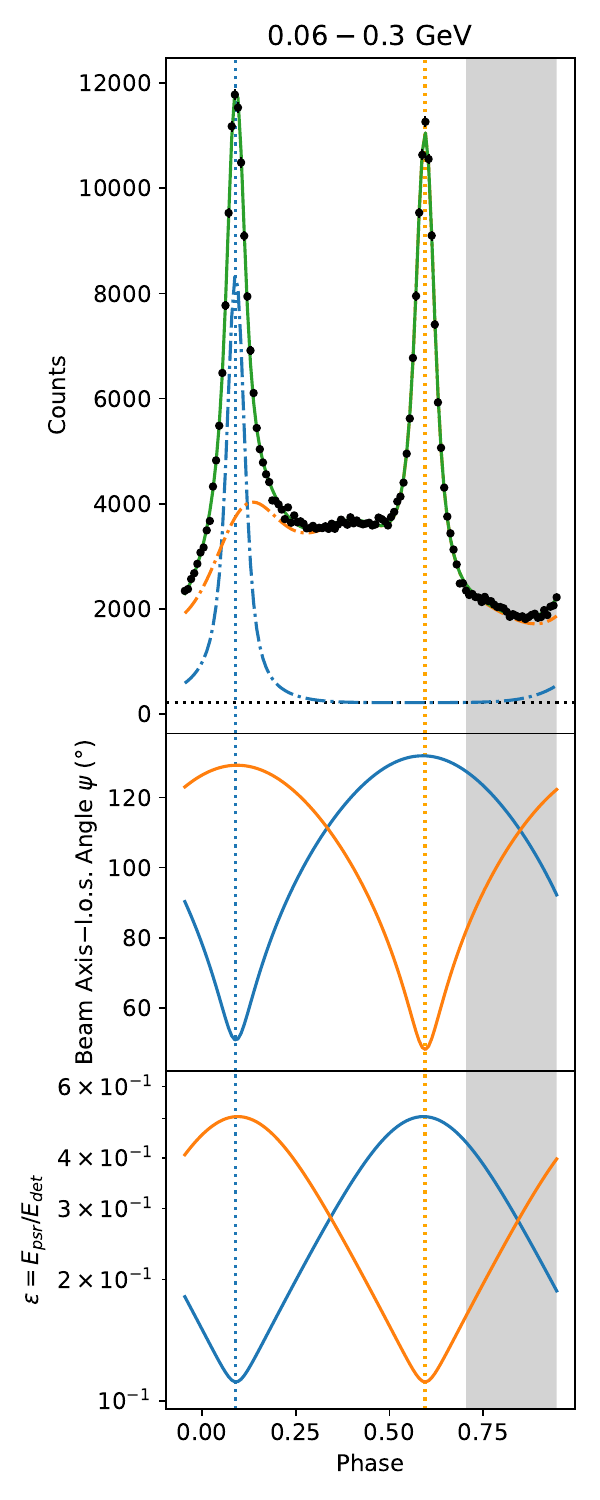}
   \includegraphics[width=.33\textwidth]{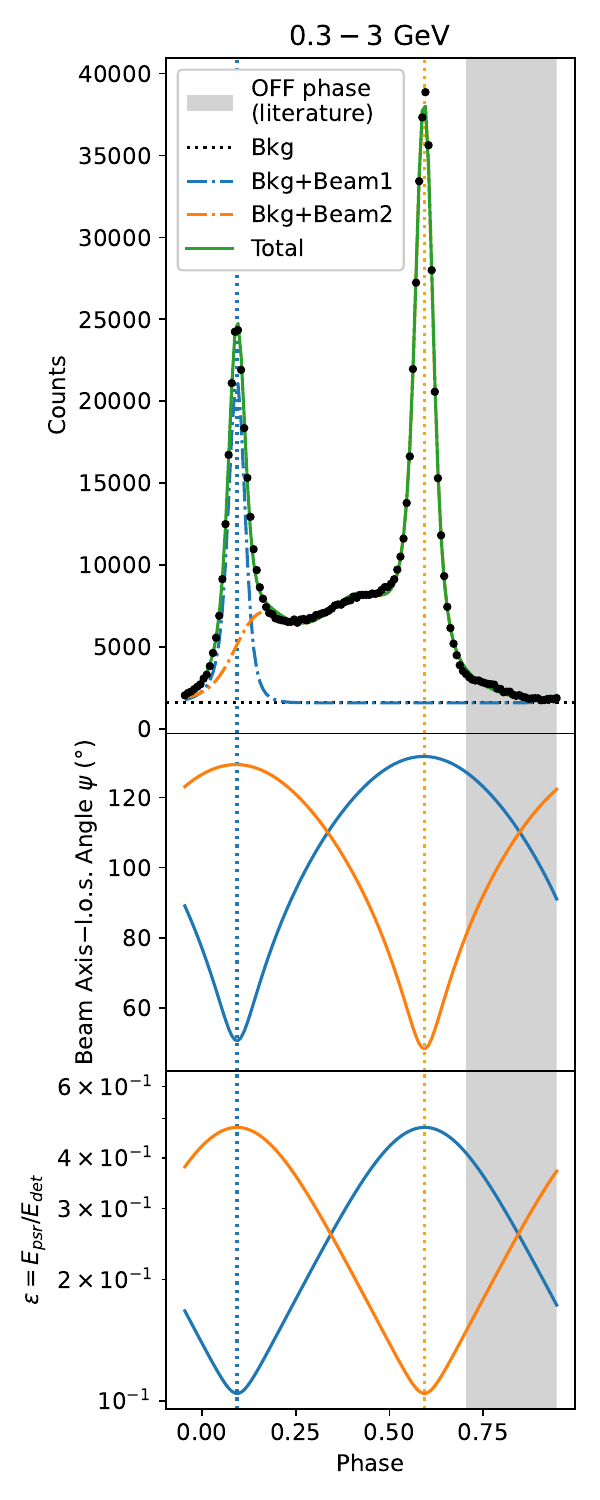}
   \includegraphics[width=.33\textwidth]{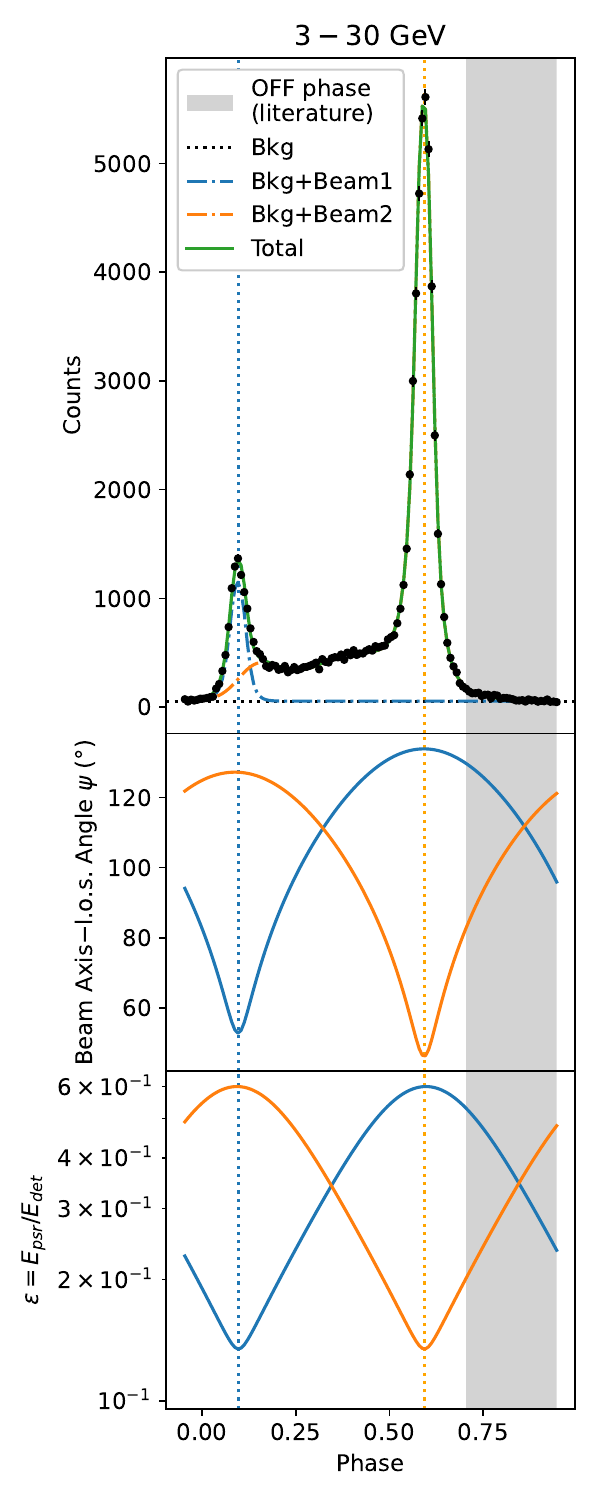}
   \caption{Crosschecking results of the Geminga pulsar, where the phaseograms are fitted with the less preferable model of PLSEC $\psi$-dependence. The panels, colours and line styles have the same meanings as in Fig.~\ref{CrabLC}.
}
              \label{Geminga_PLSEC}%
    \end{figure*}

\begin{figure*}
   \centering
   \includegraphics[width=.33\textwidth]{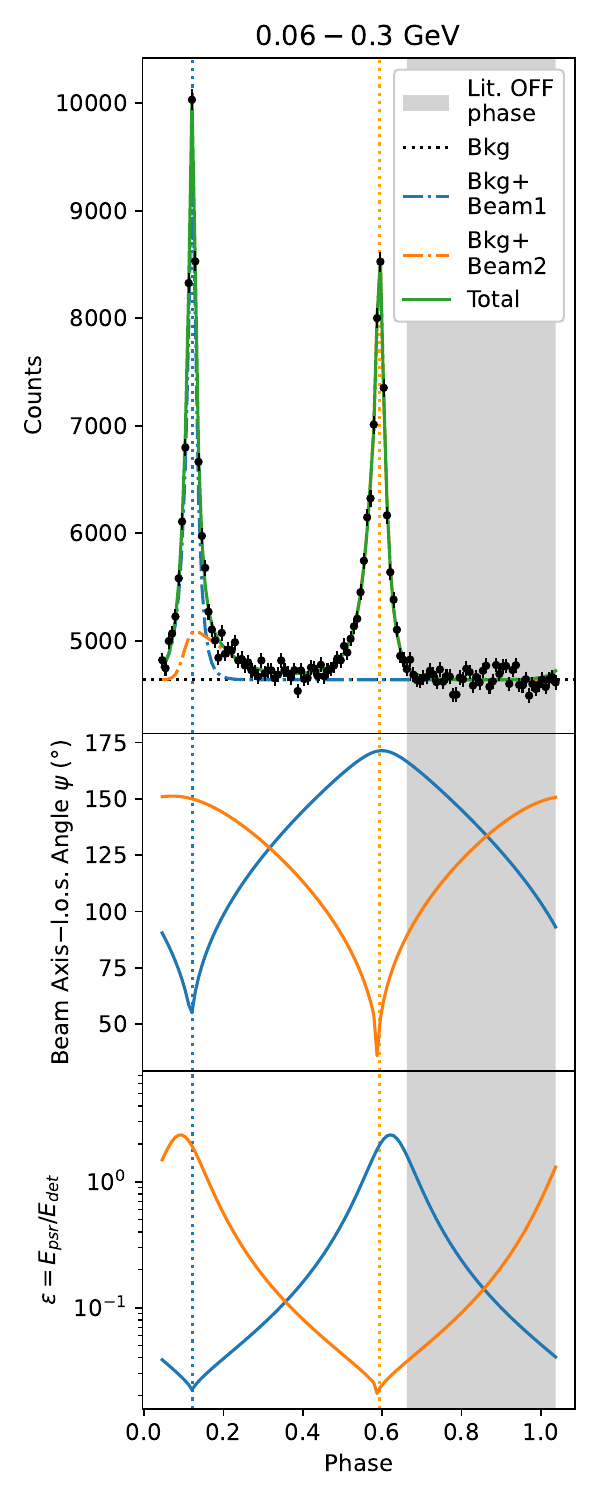}
   \includegraphics[width=.33\textwidth]{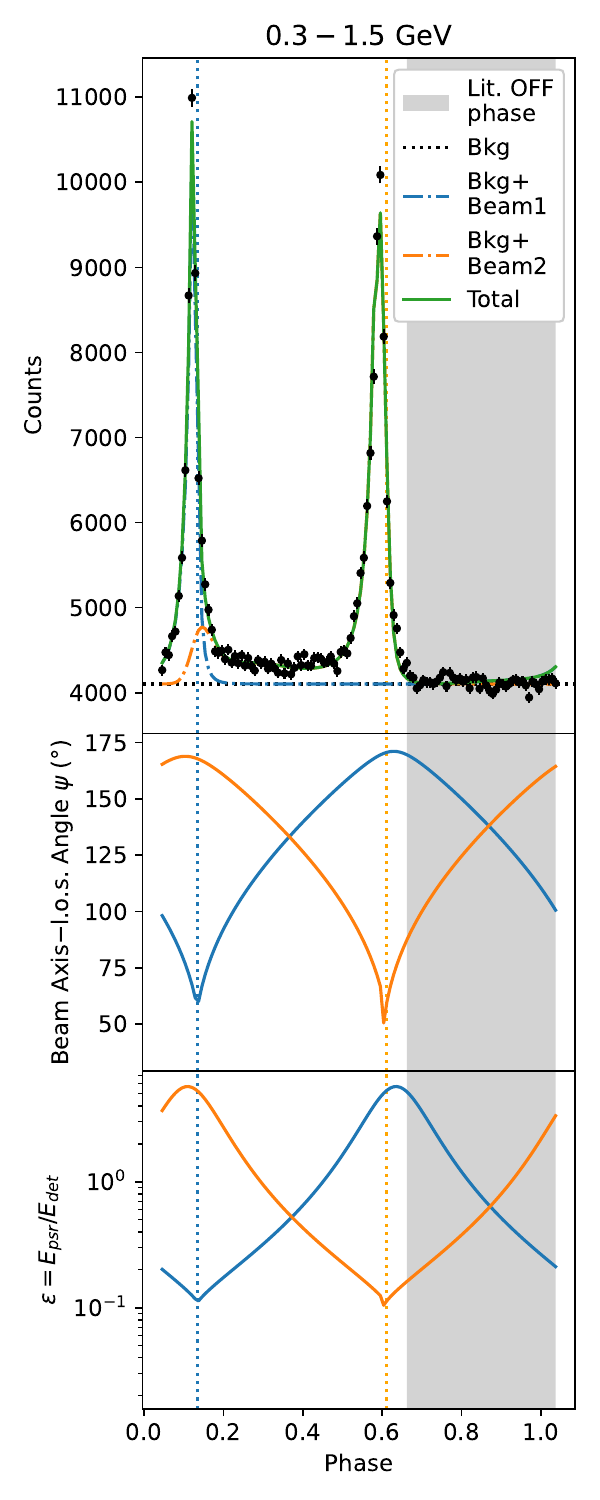}
   \includegraphics[width=.33\textwidth]{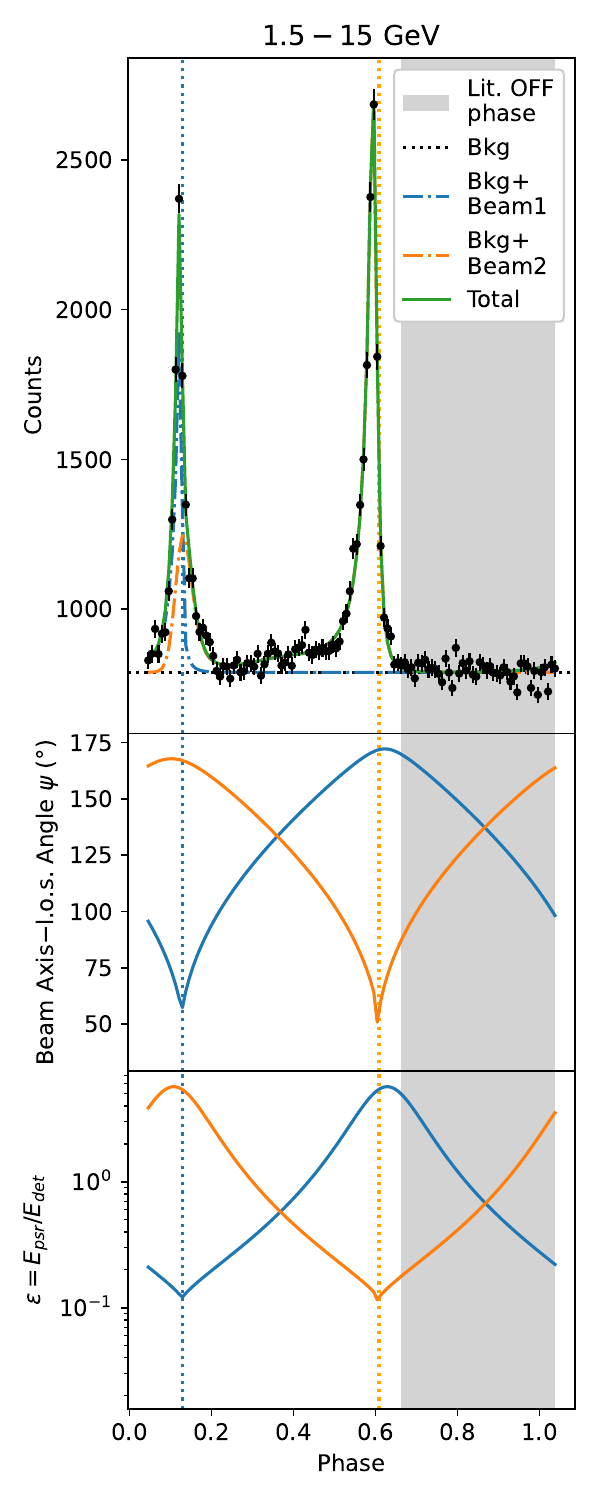}
   \caption{Crosschecking results of the Dragonfly pulsar, where the phaseograms are fitted with the less preferable model of PLSEC $\psi$-dependence. The panels, colours and line styles have the same meanings as in Fig.~\ref{CrabLC}.
}
              \label{Dragonfly_PLSEC}%
    \end{figure*}

\begin{figure}
   \centering
   \includegraphics[width=.495\columnwidth]{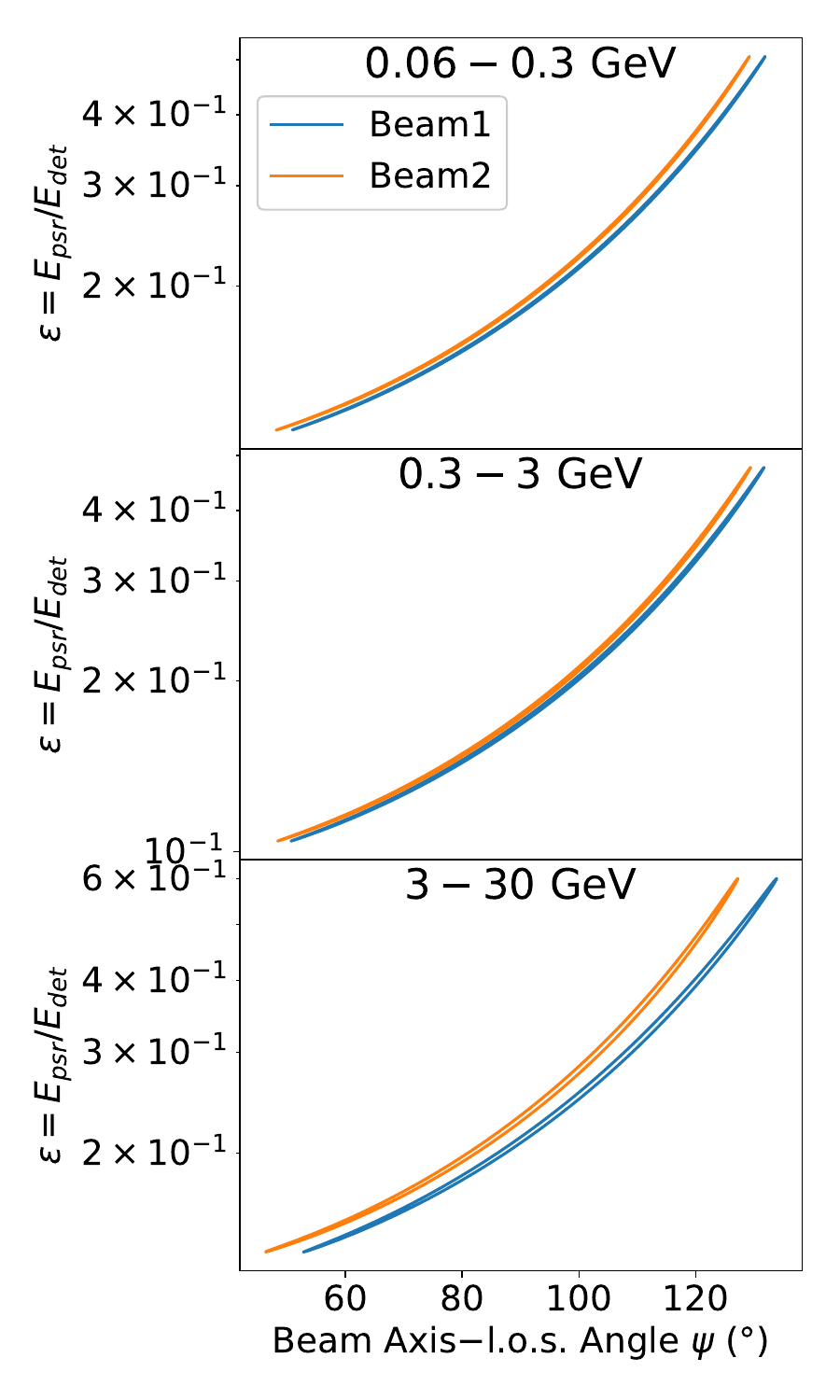}
   \includegraphics[width=.495\columnwidth]{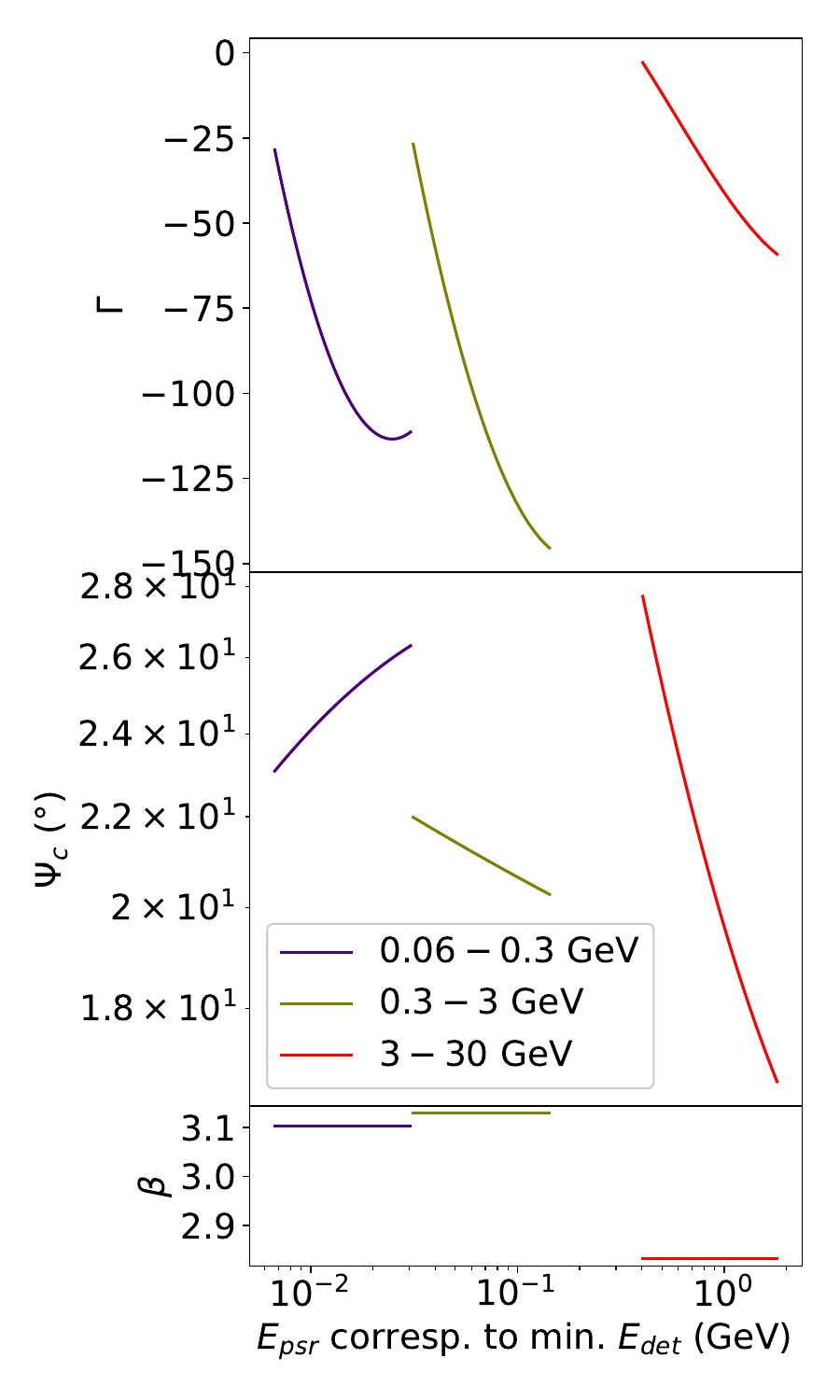}
   \caption{Systematic evaluations for the crosschecking results of the Geminga pulsar, where the less preferable model of PLSEC $\psi$-dependence is fitted. The structure of this figure is the same as in Fig.~\ref{Crab_check}.
}
              \label{Geminga_evaluate_ALT}%
    \end{figure}

\begin{figure}
   \centering
   \includegraphics[width=.495\columnwidth]{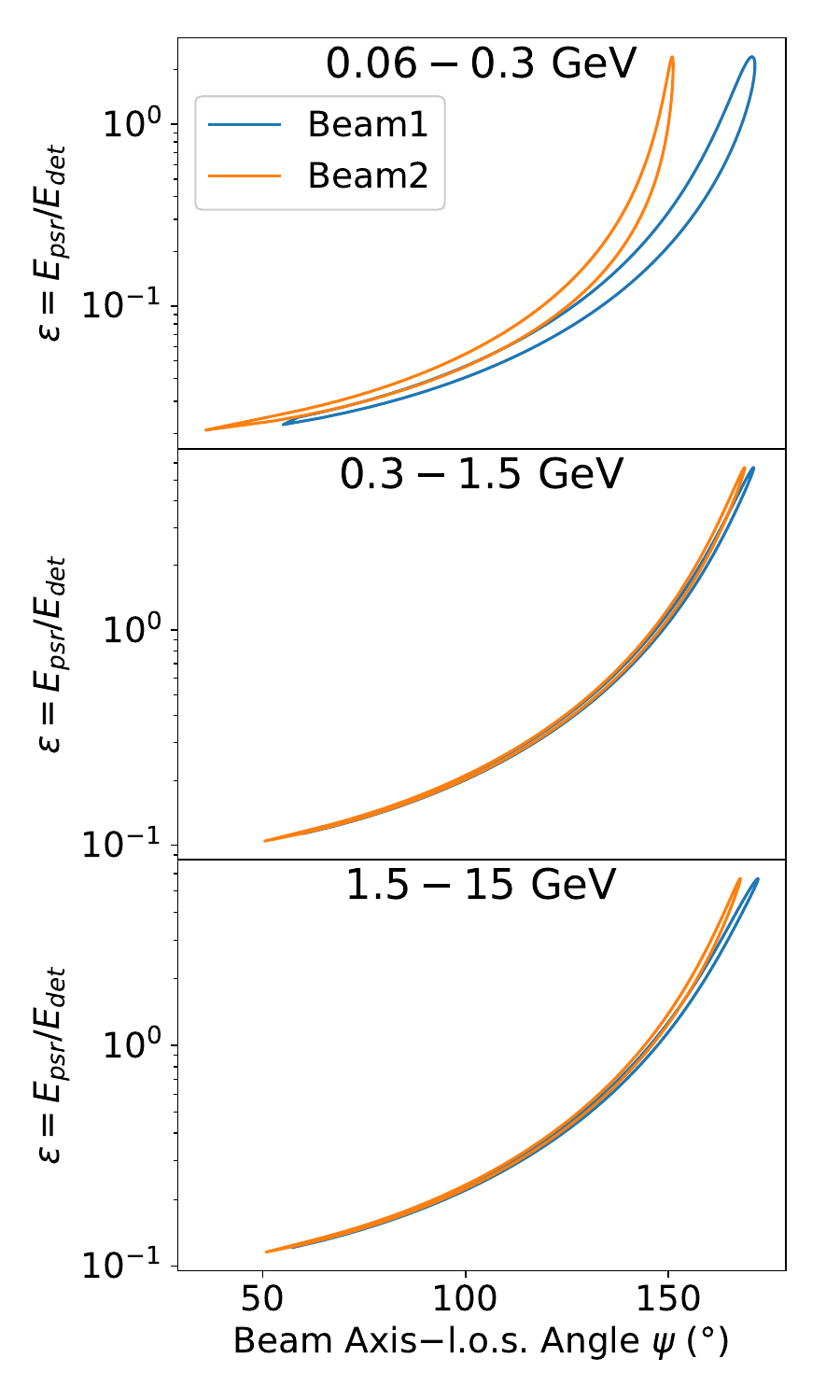}
   \includegraphics[width=.495\columnwidth]{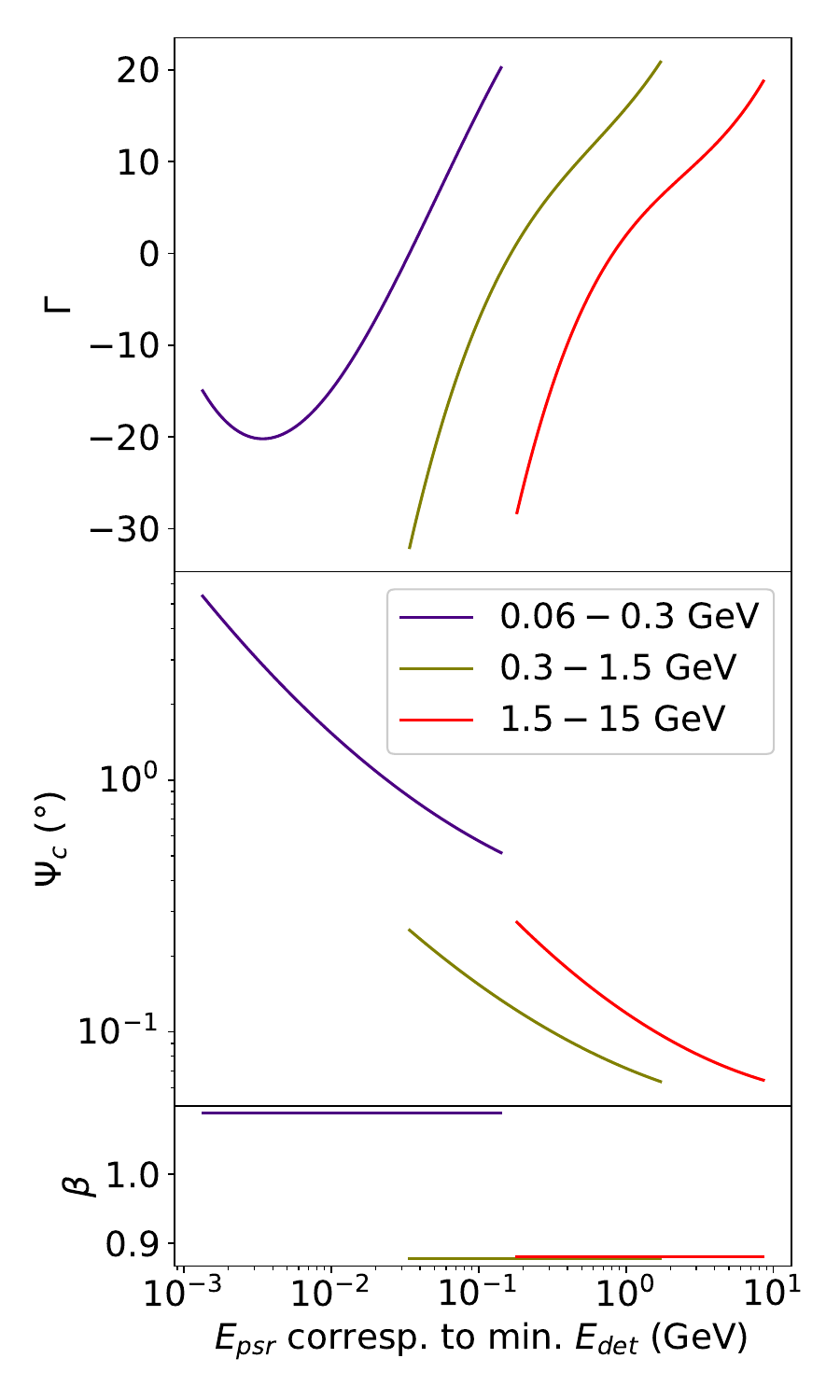}
   \caption{Systematic evaluations for the crosschecking results of the Dragonfly pulsar, where the less preferable model of PLSEC $\psi$-dependence is fitted. The structure of this figure is the same as in Fig.~\ref{Crab_check}.
}
              \label{Dragonfly_evaluate_ALT}%
    \end{figure}

As a crosscheck, we also look into the less preferable models of PLSEC $\psi$-dependence  for the Geminga and Dragonfly pulsars (Fig.~\ref{Geminga_PLSEC}~\&~\ref{Dragonfly_PLSEC}). The corresponding systematic evaluations are presented in Fig.~\ref{Geminga_evaluate_ALT}~\&~\ref{Dragonfly_evaluate_ALT}.

\subsection{Geminga pulsar}

Consistently, the less preferable model for the Geminga pulsar also predicts that the un-pulsed count rate in 0.06$-$0.3~GeV is much lower than the detected count rate at any phase, and the un-pulsed count rate in 0.3$-$3~GeV is $\sim$91\% of the minimum detected count rate. Both the models of PLSEC $\psi$-dependence and MLP $\psi$-dependence suggest that, at lower $\gamma$-ray energies, the pulsed emission of the Geminga pulsar spans over its full phase, putting the accuracy of the traditional pulsar gating technique in question.

Noticeably, PLSEC $\psi$-dependence predicts the sub-spike features of the Geminga pulsar to be less obvious than the predictions by MLP $\psi$-dependence. This is an indication that the pulse shape inferred by our methodology is moderately dependent on the assumed beam morphological model.

The model of PLSEC $\psi$-dependence for the Geminga pulsar yields $\Gamma$ that is always negative and generally decreasing with $\varepsilon$ (Figure~\ref{Geminga_evaluate_ALT}). This further strengthens the claim that the beam emitted from each hemisphere of the Geminga pulsar has a hollow conic structure. Unlike what we found for Crab and Vela pulsars,  relatively mild super-exponential cutoffs ($2.8<\beta<3.2$) are implied for the Geminga pulsar. $\theta_B$ and $\theta_C$ yielded by the model of PLSEC $\psi$-dependence are approximately straight angles too, further supporting a scenario of emission sites almost on the equatorial plane and almost horizontal beam axes. 

\subsection{Dragonfly pulsar}

\PKHY{The less preferable model of PLSEC $\psi$-dependence  for the Dragonfly pulsar strengthens the arguments regarding hollow conic beams (Fig.~\ref{Dragonfly_evaluate_ALT}) and almost horizontal beam axes. It also gives relatively large values of $\theta_B$ ($65.0\degree$--$72.8\degree$), supporting that the emission sites are fairly close to the equatorial plane. Uniquely, the predicted cutoffs of PLSEC beam shapes for the Dragonfly pulsar are approximate to ordinary exponentials ($\beta\sim1$).}

\section{Relation of a detected count rate with a source-frame intensity}\label{ApxIntensity}

Our prototypical toy model formulates the detected count rate of a beam ($C$) as a function of beam axis--l.o.s. angle ($\psi$) and inverse Doppler factor ($\varepsilon=E_{psr}/E_{det}$). Meanwhile, it is important to link $C$ with the beam’s source-frame intensity. We make the following derivations with reference to \citet{Johnson1982}.

Each of Beam1 and Beam2 considered in this work refers to a macroscopic collection of countless microscopic rays emitted by countless particles. Each microscopic ray contributes a differential count rate ($dC$) detected at a certain $E_{det}$ along a certain direction. As a result of Doppler shift, $dC$ is changed by a factor of 1/$\varepsilon$ with respect to the pulsar frame. Meanwhile, each differential effective solid angle of a detector ($d\Omega_{det}$ that each photon propagates into) is changed by a factor of $\varepsilon^2$. The detected “specific intensity” ($I_{det}$; the detected power per unit area per unit solid angle per unit frequency interval) is proportional to $\frac{E_{det}dC}{dE_{det}d\Omega_{det}}$. Eqn.~\ref{DSint} is derived from these concepts.

We stress that $E_{det}$ and $\Omega_{det}$ are instrumental properties and phase-independent. Thus, $I_{det}$ is conceptually equivalent to $C$ multiplied by some proportional constants. In turn, the emitted specific intensity of a beam in the pulsar frame ($I_{psr}$) is proportional to $\varepsilon^3{C}$, where the power of 3 is dimensionally homogeneous to $\eta_1$ of Eqn.~\ref{N_delta}.

\section{Emission site beyond a light cylinder}\label{ApxBeyondLC}

For some phaseograms, we determine the most representative location of the emission site to be beyond the light cylinder ($d\sin{\theta_B}>R_{LC}$), as shown in Tables~\ref{PLSEC_values}~\&~\ref{MLP_values}. In such cases, the cluster of its constituent particles is drifting incoherently with the pulsar’s rotation. On the other hand, the photon-emitting particles are constantly replaced in a wave-like manner. In other words, the population of particles constituting the emission region at one phase is different from (but overlapping with) that at another phase. Taking this effect into account, the "virtual motion" of the emission region is unrestricted by the speed of light and can synchronise with the pulsar’s rotation. This is the reason why an emission site beyond a light cylinder can still generate a pulse signal that we detect. A detailed astrophysical theory related to this aspect can be found in \citet{Aharonian_wind_2012}, for example.

In such a scenario, there is a segregation between the pulsar's co-rotating frame and the pulsar's emission site frame. With regards to this, we clarify that the inverse Doppler factor ($\varepsilon=E_{psr}/E_{det}$) is actually referenced to the pulsar's emission site frame (i.e. a frame where the average drift velocity of the emission site is 0 but the pulsar is still rotating).


\bsp	
\label{lastpage}
\end{document}